\documentclass[twocolumn,showpacs,preprintnumbers,amsmath,amssymb]{revtex4-1}
\usepackage{epsfig}
\usepackage{graphicx}
\usepackage{epsfig}
\usepackage{amssymb}
\usepackage{amsmath}
\usepackage{graphicx}
\usepackage{float}

\begin{document}
\title{Improved contact prediction in proteins: \\ Using
  pseudolikelihoods to infer Potts models} \author{ Magnus
  Ekeberg$^1$, Cecilia L\"{o}vkvist$^3$,Yueheng Lan$^5$, Martin
  Weigt$^6$, Erik Aurell$^{2,3,4,\dagger}$}\thanks{$^1$ Engineering
  Physics Program, KTH Royal Institute of Technology, 100 44
  Stockholm, Sweden, $^2$ ACCESS Linnaeus Center, KTH, Sweden, $^3$
  Department of Computational Biology, AlbaNova University Center, 106
  91 Stockholm, Sweden,$^4$ Aalto University School of Science,
  Helsinki, Finland, $^5$ Department of Physics, Tsinghua University,
  Beijing 100084, P. R. China, $^6$ Universit\'e Pierre et Marie
  Curie, UMR7238 -- Laboratoire de G\'enomique des Microorganismes, 15
  rue de l'Ecole de M\'edecine, 75006 Paris,
  France,$^\dagger$eaurell@kth.se} \date{\today} \pacs{02.50.Tt --
  Inference methods, 87.10.Vg -- Biological information, 87.15.Qt --
  Sequence analysis, 87.14.E- -- Proteins }

\begin{abstract}
Spatially proximate amino acids in a protein tend to coevolve. A
protein's 3D structure hence leaves an echo of correlations in the
evolutionary record. Reverse engineering 3D structures from such
correlations is an open problem in structural biology, pursued with
increasing vigor as more and more protein sequences continue to fill
the data banks. Within this task lies a statistical inference problem,
rooted in the following: correlation between two sites in a protein
sequence can arise from firsthand interaction, but can also be
network-propagated via intermediate sites; observed correlation is not
enough to guarantee proximity. To separate direct from indirect
interactions is an instance of the general problem of \textit{inverse
  statistical mechanics}, where the task is to learn model parameters
(fields, couplings) from observables (magnetizations, correlations,
samples) in large systems.  In the context of protein sequences, the
approach has been referred to as \textit{direct-coupling
  analysis}. Here we show that the pseudolikelihood method, applied to
21-state Potts models describing the statistical properties of
families of evolutionarily related proteins, significantly outperforms
existing approaches to the direct-coupling analysis, the latter being
based on standard mean-field techniques. This improved performance
also relies on a modified score for the coupling strength. The results
are verified using known crystal structures of specific sequence
instances of various protein families. Code implementing the new
method can be found at \url{http://plmdca.csc.kth.se/}.
\end{abstract}
\maketitle

\section{Introduction}
\label{sec:intro}

In biology, new and refined experimental techniques have triggered a
rapid increase in data availability during the last few years. Such
progress needs to be accompanied by the development of appropriate
statistical tools to treat growing data sets. An example of a branch
undergoing intense growth in the amount of existing data is
\textit{protein structure prediction} (PSP), which, due to the strong
relation between a protein's structure and its function, is one
central topic in biology. As we shall see, one can accurately estimate
the 3D structure of a protein by identifying which amino-acid
positions in its chain are statistically coupled over evolutionary
time
scales~\cite{Weigt,pmid22106262,journals/corr/abs-1110-5091,Sulkowska2012,Nugent}.
Much of the experimental output is today readily accessible through
public databases such as Pfam
\cite{journals/nar/PuntaCEMTBPFCCHHSEBF12}, which collects over 13,000
families of evolutionarily related protein domains, the largest of
them containing more than $2\times 10^5$ different amino-acid
sequences. Such databases allow researchers to easily access data, to
extract information from it, and to confront their results.

A recurring difficulty when dealing with interacting systems is
distinguishing direct interactions from interactions mediated via
multi-step paths across other elements. Correlations are, in general,
straightforward to compute from raw data, whereas parameters
describing the true causal ties are not. The network of direct
interactions can be thought of as hidden beneath observable
correlations, and untwisting it is a task of inherent intricacy. In
PSP, using mathematical means to dispose of the network-mediated
correlations observable in alignments of evolutionarily related (and
structurally conserved) proteins is necessary to get optimal results
\cite{Weigt,journals/ploscb/BurgerN10,Balakrishnan,pmid22106262,journals/bioinformatics/JonesBCP12}
and yields improvements worth the computational strain put on the
analysis. This approach to PSP, which we refer to as
\textit{direct-coupling analysis} (DCA), is the focus of this paper.

In a more general setting, the problem of inferring interactions from
observations of instances amounts to \textit{inverse statistical
  mechanics}, a field which has been intensively pursued in
statistical physics over the last
decade~\cite{Kappen98boltzmannmachine,Schneidman06,PhysRevLett.96.030201,1742-5468-2008-12-P12001,Frontiers,CoccoLeiblerMonasson,SessakMonasson,MezardMora,Marinari,CoccoMonasson2011,Ricci-Tersenghi2012,Nguyen-Berg2012a,Nguyen-Berg2012b,1107.3536v2}. Similar
tasks were formulated earlier in statistics and machine learning, where
they have been called \textit{model learning} and
\textit{inference}~\cite{hyvarinen2001independent,rissanen2007,WainwrightJordan,RavikumarWainwrightLafferty10}.
To illustrate this concretely, let us start from an Ising model,
\begin{equation}
\label{eq:Ising}
P(\sigma_1,\ldots,\sigma_N) = \frac{1}{Z}\exp{\left(\sum_{i=1}^N h_i\sigma_i 
+ \sum_{1\leq i<j\leq N}J_{ij}\sigma_i\sigma_j\right)}
\end{equation}
and its magnetizations $m_i =\partial_{h_i}\log Z$ and connected
correlations $c_{ij}=\partial_{J_{ij}}\log Z-m_im_j$.  Counting the
number of observables ($m_i$ and $c_{ij}$) and the number of
parameters ($h_i$ and $J_{ij}$) it is clear that perfect knowledge of
the magnetizations and correlations should suffice to determine the
external fields and the couplings exactly.  It is, however, also clear
that such a process must be computationally expensive, since it
requires the computation of the partition function $Z$ for an
arbitrary set of parameters.  The exact but iterative procedure known
as Boltzmann machines~\cite{Ackley85alearning} does in fact work on
small systems, but it is out of question for the problem sizes
considered in this paper. On the other hand, the mean-field
equations of (\ref{eq:Ising}) read \cite{Parisi,Peliti,Fischer-Hertz}:
\begin{equation}
\label{eq:Ising-mean-field}
\tanh^{-1}m_i = h_i + \sum_j J_{ij} m_j.
\end{equation}
From (\ref{eq:Ising-mean-field}) and the fluctuation-dissipation relations 
an equation can be derived connecting the coupling 
coefficients $J_ {ij}$ and the correlation matrix
${\bf c}=(c_{ij})$~\cite{Kappen98boltzmannmachine}:
\begin{equation}
J_{ij} = -({\bf c}^{-1})_{ij}.
\label{eq:Ising-mean-field-inverse}
\end{equation}
Equations~(\ref{eq:Ising-mean-field}) and
(\ref{eq:Ising-mean-field-inverse}) exemplify typical aspects of
inverse statistical mechanics, and inference in large systems in
general.  On one hand, the parameter reconstruction using these two
equations is \textit{not exact}. It is only \textit{approximate},
because the mean-field equations (\ref{eq:Ising-mean-field}) are
themselves only approximate. It also demands reasonably good sampling,
as the matrix of correlations is not invertible unless it is of full
rank, and small noise on its ${\cal O}(N^2)$ entries may result in
large errors in estimating the $J_{ij}$.  On the other hand, this
method is \textit{fast}, as fast as inverting a matrix, because one
does not need to compute $Z$.  Except for mean-field methods as in
(\ref{eq:Ising-mean-field}), approximate methods recently used to
solve the inverse Ising problem can be grouped as \textit{expansion in
  correlations and clusters} \cite{SessakMonasson,CoccoMonasson2011},
methods based on the \textit{Bethe approximation}
\cite{MezardMora,Marinari,Ricci-Tersenghi2012,Nguyen-Berg2012a,Nguyen-Berg2012b},
and the \textit{pseudolikelihood method}
\cite{RavikumarWainwrightLafferty10,1107.3536v2}.

For PSP, it is not the Ising model but a 21-state Potts model which is
pertinent~\cite{Weigt}: The model shall be learned such that it
represents the statistical features of large multiple sequence
alignments of homologous (evolutionarily related) proteins, and to
reproduce the statistics of correlated amino acid substitutions. This
can be done with the Potts equivalent of Eq.~(\ref{eq:Ising}), {\it
  i.e.} using a model with pairwise interactions. As will be detailed
below, strong interactions can be interpreted as indicators for
spatial vicinity of residues in the three-dimensional protein fold,
even if residues are well separated along the sequence -- thus linking
evolutionary sequence statistics with protein structure. But which of
all the inference methods in inverse statistical mechanics, machine
learning or statistics is most suitable for treating real protein
sequence data? How do the test results obtained for independently
generated equilibrium configurations of Potts models translate to the
case of protein sequences, which are neither independent nor
equilibrium configurations of any well-defined statistical-physics
model? The main goal of this paper is to move towards an answer to
this question by showing that the pseudolikelihood method is a very
powerful means to perform DCA, with a prediction accuracy considerably
out-performing methods previously assessed.

The paper is structured as follows: in Section~\ref{sec:psp}, we
review the ideas underlying PSP and DCA and explain the biological
hypotheses linking protein 3D structure to correlations in amino-acid
sequences. We also review earlier approaches to DCA.  In
Section~\ref{sec:method}, we describe the Potts model in the context
of DCA and the properties of exponential families. We further detail a
maximum-likelihood (ML) approach as brought to bear on the inverse
Potts problem and discuss in more detail why this is impractical for
realistic system sizes, and we introduce, similarly to
(\ref{eq:Ising-mean-field-inverse}) above, the inverse Potts
mean-field model algorithm for the DCA (mfDCA) and a pseudolikelihood
maximization procedure (plmDCA).  This section also covers algorithmic
details of both models such as regularization and sequence
reweighting. A further important issue is the selection of an
interaction score, which reduces coupling matrices to a scalar score,
and allows for ranking of couplings according to their 'strength'.
In Section~\ref{sec:exp}, we present results from prediction
experiments using mfDCA and plmDCA assessed against known crystal
structures. In Section~\ref{sec:discussion}, we summarize our
findings, put their implications into context, and discuss possible
future developments. The appendixes contain additional material supporting
the main text.

\section{Protein Structure Prediction and Direct-Coupling Analysis}
\label{sec:psp}

Proteins are essential players in almost all biological
processes. Primarily, proteins are linear chains, each site being
occupied by 1 out of 20 different amino acids. Their function
relies, however, on the protein \textit{fold}, which refers to the 3D
conformation into which the amino-acid chain curls. This fold
guarantees, e.g., that the right amino acids are exposed in the right
positions at the protein surface, thus forming biochemically active
sites, or that the correct pairs of amino acids are brought into
contact to keep the fold thermodynamically stable.

Experimentally determining the fold, using e.g. x-ray
crystallography or NMR tomography, is still a rather costly and
time-consuming process. On the other hand, every newly sequenced
genome results in a large number of newly predicted proteins. The
number of sequenced organisms now exceeds $3,700$, and
continues to grow exponentially ({\tt genomesonline.org}
\cite{GOLD}). The most prominent database for protein sequences,
Uniprot ({\tt uniprot.org} \cite{Uniprot}), lists about 25,000,000
different protein sequences, whereas the number of experimentally
determined protein structures is only around 85,000 ({\tt pdb.org}
\cite{pdb}).

However, the situation of structural biology is not as hopeless as
these numbers might suggest. First, proteins have a modular
architecture; they can be subdivided into {\it domains} which, to a
certain extent, fold and evolve as units. Second, domains of a common
evolutionary origin, i.e., so-called {\it homologous} domains, are
expected to almost share their 3D structure and to have related
function. They can therefore be collected in {\it protein domain
  families}: the Pfam database ({\tt pfam.sanger.ac.uk}
\cite{journals/nar/PuntaCEMTBPFCCHHSEBF12}) lists almost 14,000
different domain families, and the number of sequences collected in
each single family ranges roughly from $10^2$ to $10^5$. In particular the larger
families, with more than 1,000 members, are of interest to us, as we argue that their natural sequence variability contains important
statistical information about the 3D structure of its member proteins,
and can be exploited to successfully address the PSP problem.

\begin{figure}[t]
  \begin{center}
    \includegraphics[width=8cm,height=4.5cm]{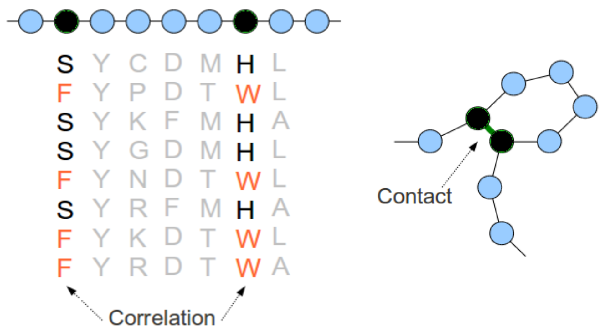}
  \end{center}
  \caption{(Color online) Left panel: small MSA with two positions of correlated
    amino-acid occupancy.  Right panel: hypothetical corresponding
    spatial conformation, bringing the two correlated positions into
    direct contact.}
  \label{fig:MSA_contact}
\end{figure}

Two types of data accessible via the Pfam database are especially
important to us. The first is the \textit{multiple sequence alignment}
(MSA), a table of the amino acid sequences of all the protein domains
in the family lined up to be as similar as possible.  A (small and
illustrative) example is shown in Fig.~\ref{fig:MSA_contact} (left
panel).  Normally, not all members of a family can be lined up
perfectly, and the alignment therefore contains both amino acids and
\textit{gaps}.  At some positions, an alignment will be highly
specific (cf.~the second, fully conserved column in
Fig.~\ref{fig:MSA_contact}), while at others it will be more variable.
The second data type concerns the \textit{crystal structure} of one or
several members of a family. Not all families provide this second type
of data. We discuss its use for an {\it a posteriori} assessment
of our inference results in detail in Sec.~\ref{sec:exp}.

The Potts-model based inference uses only the first data type,
i.e., the sequence data. Small spatial separation between amino acids in a
protein, cf. the right panel of Fig.~\ref{fig:MSA_contact}, encourages
co-occurrence of favorable amino-acid combinations, cf. the left
panel of Fig.~\ref{fig:MSA_contact}. This spices the sequence record
with correlations, which can be reliably determined in sufficiently
large MSAs. However, the use of such correlations for predicting 3D
contacts as a first step to solve the PSP problem remained of limited
success \cite{pmid8208723,Lockless08101999,PROT:PROT20098}, since they
can be induced both by direct interactions (amino acid $A$ is close to
amino acid $B$), and by indirect interactions (amino acids $A$
and $B$ are both close to amino acid $C$). Lapedes \textit{et
  al.}~\cite{Lapedes} were the first to address, in a purely
theoretical setting, these ambiguities of a correlation-based route to
protein sequence analysis, and these authors also outline a
maximum-entropy approach to get at direct interactions.
Weigt~\textit{et al.} \cite{Weigt} successfully executed this program
subsequently called \textit{direct-coupling analysis}: the accuracy in
predicting contacts strongly increases when direct interactions are
used instead of raw correlations.

To computationally solve the task of inferring interactions in a Potts
model, \cite{Weigt} employed a generalization of the iterative
message-passing algorithm \emph{susceptibility propagation} previously
developed for the inverse Ising problem~\cite{MezardMora}.  Methods in
this class are expected to outperform mean-field based reconstruction
methods similar to (\ref{eq:Ising-mean-field-inverse}) if the
underlying graph of direct interactions is locally close to tree-like,
an assumption which may or may not be true in a given application such
as PSP. A substantial draw-back of susceptibility propagation as used
in~\cite{Weigt} is that it requires a rather large amount of auxiliary
variables, and that DCA could therefore only be carried out on protein sequences that are not too
long. In \cite{pmid22106262}, this obstacle was
overcome by using instead a simpler mean-field method,
i.e., the generalization of
(\ref{eq:Ising-mean-field-inverse}) to a 21-state Potts model. As
discussed in \cite{pmid22106262}, this broadens the reach of the DCA
to practically all families currently in Pfam. It improves the
computational speed by a factor of about $10^3$--$10^4$, and it
appears also to be more accurate than the susceptibility-propagation based
method of \cite{Weigt} in predicting contact pairs.  The reason behind
this third advantage of mean-field over susceptibility propagation as
an approximate method of DCA is unknown at this time.

\textit{Pseudolikelihood maximization} (PLM) is an alternative method
developed in mathematical statistics to approximate maximum-likelihood
inference, which breaks down the {\it a priori} exponential
time complexity of computing partition functions in 
exponential families~\cite{besag1975}. On the inverse Ising problem it was first
used by Ravikumar \textit{et al.}~\cite{RavikumarWainwrightLafferty10},
albeit in the context of graph sign-sparsity reconstruction; two of us showed
recently that it outperforms many other approximate inverse Ising
schemes on the Sherrington-Kirkpatrick model, and in several other
examples~\cite{1107.3536v2}. Although this paper reports the first use of
the PLM method in DCA, the idea of using
pseudolikelihoods for PSP is not completely novel. Balakrishnan {\it
  et al.} \cite{Balakrishnan} devised a version of this idea, but
using a set-up rather different from that in
\cite{pmid22106262}, regarding, for example, what portions of the data
bases and which measures of prediction accuracy were used, and
also, not couched in the language of inverse statistical mechanics.  While a
competitive evaluation between \cite{pmid22106262} and
\cite{Balakrishnan} is still open, we have not attempted such a comparison in
this work.

Other ways of deducing direct interactions in PSP, not motivated from
the Potts model but in somewhat similar probabilistic settings, have been
suggested in the last few years. A fast method utilizing Bayesian
networks was provided by Burger and van Nimwegen
\cite{journals/ploscb/BurgerN10}. More recently, Jones {\it et al.}
\cite{journals/bioinformatics/JonesBCP12} introduced a procedure
called \emph{PSICOV} (Protein Sparse Inverse COVariance). While DCA
and PSICOV both appear capable of outperforming the Bayesian network
approach \cite{pmid22106262,journals/bioinformatics/JonesBCP12}, their
relative efficiency is currently open to investigation, and is not assessed in this work.

Finally, predicting amino acid contacts is not only a goal in itself, but also a means to assemble protein
complexes~\cite{Schug2009,Dago2012} and to predict full 3D protein
structures~\cite{journals/corr/abs-1110-5091,Sulkowska2012,Marks2012-Cell}.
Such tasks require additional work, using the DCA results as input,
and are outside the scope of the present paper.

\section{Method development}
\label{sec:method}

\subsection{The Potts model}
Let $\boldsymbol{\sigma}=(\sigma_1,\sigma_2,\cdots,\sigma_N)$
represent the amino acid sequence of a domain with length $N$. Each
$\sigma_i$ takes on values in $\{1, 2, ..., q\}$, with $q=21$: one
state for each of the 20 naturally occurring amino acids and one
additional state to represent gaps. Thus, an MSA with $B$ aligned
sequences from a domain family can be written as an integer array
$\{\boldsymbol{\sigma}^{(b)}\}^B_{b=1}$, with one row per sequence and
one column per chain position. Given an MSA, the empirical individual
and pairwise \textit{frequencies} can be calculated as
\begin{eqnarray}
\label{frequencies}
f_i(k)
&=&\frac{1}{B} \sum^B_{b=1} \delta(\sigma^{(b)}_i,k), \nonumber\\
f_{ij}(k,l)
&=&\frac{1}{B} \sum^B_{b=1} \delta(\sigma^{(b)}_i,k)\ 
\delta(\sigma^{(b)}_j,l),
\end{eqnarray}
where $\delta(a,b)$ is the Kronecker symbol taking value 1 if both
arguments are equal, and 0 otherwise. $f_i(k)$ is hence the fraction of
sequences for which the entry on position $i$ is amino acid $k$, gaps
counted as a 21st amino acid. Similarly, $f_{ij}(k,l)$ is the fraction
of sequences in which the position pair $(i,j)$ holds the amino acid
combination $(k,l)$. Connected correlations are given as
\begin{equation}
\label{correlations}
c_{ij}(k,l)=f_{ij}(k,l) - f_{i}(k)\,f_{j}(l).
\end{equation}

A (generalized) Potts model is the simplest probabilistic model
$P(\boldsymbol{\sigma})$ which can reproduce the empirically observed
$f_i(k)$ and $f_{ij}(k,l)$. In analogy to (\ref{eq:Ising}) it is
defined as
\begin{equation}
\label{potts_model}
P(\boldsymbol{\sigma})= \frac{1}{Z} \exp\left(\sum\limits^N_{i=1} 
{h_i(\sigma_i)}+\sum_{1\leq i<j\leq N}{ J_{ij}(\sigma_i,\sigma_j)}\right),
\end{equation}
in which $h_i(\sigma_i)$ and $J_{ij}(\sigma_i,\sigma_j)$ are
parameters to be determined through the constraints
\begin{eqnarray}
\label{maxent_constraints}
P(\sigma_i=k) & = & \sum\limits_{\underset{\sigma_i=k}{\boldsymbol{\sigma}}} 
P(\boldsymbol{\sigma})= f_i(k), \nonumber \\
P(\sigma_i=k,\sigma_j=l) & = & 
\sum\limits_{\underset{\underset{\sigma_i=k}{\sigma_j=l}}{\boldsymbol{\sigma}}} 
P(\boldsymbol{\sigma})= f_{ij}(k,l).
\end{eqnarray}
It is immediate that the probabilistic model which maximizes the
entropy while satisfying Eq.~(\ref{maxent_constraints}) must take the
Potts model form. Finding a Potts model which matches empirical
frequencies and correlations is therefore referred to as a
\textit{maximum-entropy inference} \cite{jaynes57a,jaynes57b},
cf. also \cite{Lapedes,Weigt} for a formulation in terms of
protein-sequence modeling.

On the other hand, we can use Eq.~(\ref{potts_model}) as a variational
ansatz and maximize the probability of the input data set
$\{\boldsymbol{\sigma}^{(b)}\}^B_{b=1}$ with respect to the model
parameters $h_i(\sigma)$ and $J_{ij}(\sigma,\sigma')$; this {\it
  maximum-likelihood} perspective is used in the following. We
note that the Ising and the Potts models (and most models which are
normally considered in statistical mechanics) are examples of
\textit{exponential families}, and have the property that means and
correlations are \textit{sufficient
  statistics}~\cite{darmois35,pitman-wishart36,koopman36}.  Given
unlimited computing power to determine $Z$, reconstruction can not be
done better using all the data compared to using only (empirical)
means and (empirical) correlations. It is only when one cannot compute
$Z$ exactly and has to resort to approximate methods, that using
directly all the data can bring any advantage.

\subsection{Model parameters and gauge invariance}

The total number of parameters of Eq,~(\ref{potts_model}) is
$Nq+\frac{N(N-1)}{2}q^2$, but, in fact, the model as it stands is
overparameterized in the sense that distinct parameter sets can
describe the same probability distribution. It is easy to see that the
number of nonredundant parameters is
$N(q-1)+\frac{N(N-1)}{2}(q-1)^2$, cf. an Ising model ($q=2$), which has
$\frac{N(N+1)}{2}$ parameters if written as in Eq.~(\ref{eq:Ising})
but would have $2N^2$ parameters if written in the form of
Eq.~(\ref{potts_model}).

A gauge choice for the Potts model, which eliminates the
overparametrization in a similar manner as in the Ising model (and
reduces to that case for $q=2$), is
\begin{equation}
\label{sum_gauge}
\sum_{s=1}^q J_{ij}(k,s)=\sum_{s=1}^q 
J_{ij}(s,l)=\sum_{s=1}^q h_i(s)=0,
\end{equation}
for all $i$, $j (>i)$, $k$, and $l$. Alternatively, we can choose a
gauge where one index, say $i=q$, is special, and measure all
interaction energies with respect to this value, i.e.,
\begin{equation}
\label{zero_gauge}
J_{ij}(q,l) = J_{ij}(k,q) = h_i(q)=0,
\end{equation}
for all $i$, $j (>i)$, $k$, and $l$, cf.~\cite{pmid22106262}. This gauge
choice corresponds to a lattice gas model with $q-1$ different
particle types, and a maximum occupation number one.

Using either (\ref{sum_gauge}) or (\ref{zero_gauge}) reconstruction is
well-defined, and it is straight-forward to translate results obtained
in one gauge to the other.

\subsection{The inverse Potts problem}

Given a set of independent equilibrium configurations
$\{\boldsymbol{\sigma}^{(b)}\}^B_{b=1}$ of the model
Eq.~(\ref{potts_model}), the ordinary statistical approach to
inferring parameters
$\{\boldsymbol{\mathrm{h}},\boldsymbol{\mathrm{J}}\}$ would be to let
those parameters maximize the likelihood (i.e., the probability of
generating the data set for a given set of parameters). This is
equivalent to minimizing the (rescaled) negative log-likelihood
function
\begin{equation}
\label{nll_def}
l=-\frac{1}{B}\sum\limits^B_{b=1} \log P(\boldsymbol{\sigma}^{(b)}).
\end{equation}
For the Potts model (\ref{potts_model}), this becomes
\begin{eqnarray}
\label{nll_potts}
l(\boldsymbol{\mathrm{h}},\boldsymbol{\mathrm{J}}) 
&=&  \log Z -\sum\limits^N_{i=1} \sum\limits^q_{k=1} {f_i(k) h_i(k)}  \\ 
&&-\sum_{1\leq i<j\leq N}  \sum\limits^q_{k,l=1} {f_{ij}(k,l) J_{ij}(k,l)} .
\nonumber 
\end{eqnarray}
$l$ is differentiable, so minimizing it means looking for a point at
which $\partial_{h_i(k)}l=0$ and
$\partial_{J_{ij}(k,l)}l=0$. Hence, ML
estimates will satisfy
\begin{eqnarray}
\label{ml_conditions}
\partial_{h_i(k)} \log Z- f_i(k) & = & 0, \nonumber \\
\partial_{J_{ij}(k,l)} \log Z- f_{ij}(k,l) & = & 0.
\end{eqnarray}
To achieve this minimization computationally, we need to be able to
calculate the partition function $Z$ of Eq.~(\ref{potts_model}) for
any realization of the parameters
$\{\boldsymbol{\mathrm{h}},\boldsymbol{\mathrm{J}}\}$. This problem is
computationally intractable for any reasonable system
size. Approximate minimization is essential, and we will show that
even relatively simple approximation schemes lead to accurate PSP
results.

\subsection{Naive mean-field inversion}

The mfDCA algorithm in \cite{pmid22106262} is based on the simplest
and computationally most efficient approximation, i.e., ~{\it naive
mean-field inversion} (NMFI). It starts from the proper generalization
of (\ref{eq:Ising-mean-field}), \textit{cf.}~\cite{Kholodenko1990},
and then uses linear response: The $J$'s in the lattice-gas gauge 
Eq.~(\ref{zero_gauge}) become:
\begin{equation}
\label{NMFI_potts}
J^{NMFI}_{ij}(k,l)=-({\boldsymbol{\mathrm{C}}}^{-1})_{ab},
\end{equation}
where $a=(q-1)(i-1)+k$ and $b=(q-1)(j-1)+l$. The matrix
$\boldsymbol{\mathrm{C}}$ is the $N(q-1) \times N(q-1)$ covariance
matrix assembled by joining the $N(q-1) \times N(q-1)$ values
$c_{ij}(k,l)$ as defined in Eq.~(\ref{correlations}), but leaving out
the last state $q$. In Eq.~(\ref{NMFI_potts}), $i,j\in\{1,...,N\}$ are
site indices, and $k,l$ run from $1$ to $q-1$. Under gauge
Eq.~(\ref{zero_gauge}), all the other coupling parameters are zero.
The term 'naive' has become customary in the inverse statistical
mechanics literature, often used to highlight the difference to a
Thouless-Anderson-Palmer level inversion or one based on the Bethe
approximation.  The original meaning of this term lies, as far as we
are aware, in information geometry~\cite{Tanaka2000,amari2001}.

\subsection{Pseudolikelihood maximization}

Pseudolikelihood substitutes the probability in (\ref{nll_def}) by the
conditional probability of observing one variable $\sigma_r$ given
observations of all the other variables
$\boldsymbol{\sigma}_{\backslash r}$.  That is, the starting point is
\begin{eqnarray}
\label{cond_prob2}
P(\sigma_r&=&\sigma_r^{(b)}|\boldsymbol{\sigma}_{\backslash r}
=\boldsymbol{\sigma}_{\backslash r}^{(b)})  \nonumber \\ 
&=& \frac{\exp\left(h_r(\sigma_r^{(b)})
+\sum\limits^N_{\underset{i\neq r}{i = 1}}{J_{ri}
(\sigma_r^{(b)},\sigma_i^{(b)})}\right)}{\sum_{l=1}^q 
\exp\left(h_r(l)
+\sum\limits^N_{\underset{i\neq r}{i = 1}}{J_{ri}(l,\sigma_i^{(b)})}\right)},
\end{eqnarray}
where, for notational convenience, we take $J_{ri}(l,k)$ to mean
$J_{ir}(k,l)$ when $i<r$.  Given an MSA, we can maximize the
conditional likelihood by minimizing
\begin{equation}
\label{g_def}
g_r(\boldsymbol{\mathrm{h}}_r,\boldsymbol{\mathrm{J}}_r )
=-\frac{1}{B}  \sum\limits^B_{b=1} \log 
\left[ P_{\{\boldsymbol{\mathrm{h}}_r,\boldsymbol{\mathrm{J}}_r \}}(\sigma_r=\sigma^{(b)}_r|
\boldsymbol{\sigma}_{\backslash r}=\boldsymbol{\sigma}^{(b)}_{\backslash r}) \right].
\end{equation}
Note that this only depends on $\boldsymbol{\mathrm{h}}_r$ and
$\boldsymbol{\mathrm{J}}_r=\{\boldsymbol{\mathrm{J}}_{ir}\}_{i \neq
  r}$, i.e., on the parameters featuring node $r$.  If
(\ref{g_def}) is used for all $r$ this leads to unique values for the
parameters $\boldsymbol{\mathrm{h}}_r$ but typically different predictions for
$\boldsymbol{\mathrm{J}}_{rq}$ and $\boldsymbol{\mathrm{J}}_{qr}$ (which should be the
same). Minimizing (\ref{g_def}) must therefore be supplemented by some
procedure on how to reconcile different values of
$\boldsymbol{\mathrm{J}}_{rq}$ and $\boldsymbol{\mathrm{J}}_{qr}$; one way would be to
simply use their average
$\frac 12 ({\boldsymbol{\mathrm{J}}_{rq}+\boldsymbol{\mathrm{J}}_{qr}^T})$~\cite{RavikumarWainwrightLafferty10}.

We here reconcile different $\boldsymbol{\mathrm{J}}_{rq}$ and $\boldsymbol{\mathrm{J}}_{qr}$
by minimizing an objective function adding $g_r$ for all nodes:
\begin{eqnarray}
\label{npll}
&&l_{pseudo}(\boldsymbol{\mathrm{h}},\boldsymbol{\mathrm{J}}) = 
\sum\limits^N_{r=1} g_r(\boldsymbol{\mathrm{h}}_r,\boldsymbol{\mathrm{J}}_r ) \\ 
&=& -\frac{1}{B} \sum\limits^N_{r=1}  \sum\limits^B_{b=1} \log 
\left[ P_{\{\boldsymbol{\mathrm{h}}_r,\boldsymbol{\mathrm{J}}_r \}}(\sigma_r=\sigma^{(b)}_r|
\boldsymbol{\sigma}_{\backslash r}=\boldsymbol{\sigma}^{(b)}_{\backslash r}) \right]
\nonumber.
\end{eqnarray}
Minimizers of $l_{pseudo}$ generally do not
minimize $l$; the replacement of likelihood with pseudolikelihood
alters the outcome. Note, however, that replacing $l$ by $l_{pseudo}$
resolves the computational intractability of the parameter
optimization problem: instead of depending on the full partition
function, the normalization of the conditional probability
(\ref{cond_prob2}) contains only a single summation over the $q=21$
Potts states. The intractable average over the $N-1$ conditioning spin
variables is replaced by an empirical average over the data set in
Eq.~(\ref{npll}).

\subsection{Regularization}
A Potts model describing a protein family with sequences of 50-300
amino acids requires ca. $5 \cdot 10^5$ to $2 \cdot 10^7$ parameters. At
present, few protein families are in this range in size, and
\textit{regularization} is therefore needed to avoid overfitting.  In
NMFI, the problem results in an empirical
covariance matrix which typically is not of full rank, and
Eq.~(\ref{NMFI_potts}) is not well-defined. In \cite{pmid22106262},
one of the authors therefore used the pseudocount method where
frequencies and empirical correlations are adjusted using a
regularization variable $\lambda$:
\begin{eqnarray}
\label{pseudo_frequencies}
f_i(k)&=&\frac{1}{\lambda+B} \left[
\frac{\lambda}{q}+\sum^B_{b=1} \delta(\sigma^{(b)}_i,k) \right], \\
f_{ij}(k,l)&=&\frac{1}{\lambda+B} \left[
\frac{\lambda}{q^2}+\sum^B_{b=1} \delta(\sigma^{(b)}_i,k)\ 
\delta(\sigma^{(b)}_j,l)\right]. \nonumber
\end{eqnarray}
The pseudocount is a proxy for many observations, which -- if they
existed -- would increase the rank of the correlation matrix; the
pseudocount method hence promotes invertibility of the matrix in
Eq.~(\ref{NMFI_potts}).  It was observed in \cite{pmid22106262} that for
good performance in DCA, the pseudocount parameter $\lambda$ has to be
taken fairly large, on the order of $B$.

In the PLM method, we take the standard
route of adding a penalty term to the objective function:
\begin{equation}
\label{PLM_est_reg}
\{\boldsymbol{\mathrm{h}}^{PLM},\boldsymbol{\mathrm{J}}^{PLM} \}
=\underset{\{\boldsymbol{\mathrm{h}},\boldsymbol{\mathrm{J}} \}}
{\mathrm{argmin}} \{l_{pseudo}(\boldsymbol{\mathrm{h}}^{},\boldsymbol{\mathrm{J}})
+R(\boldsymbol{\mathrm{h}},\boldsymbol{\mathrm{J}})\}.
\end{equation}
The turnout is then a trade-off between likelihood maximization and
whatever qualities $R$ is pushing for. Ravikumar {\it et
  al.}~\cite{RavikumarWainwrightLafferty10} pioneered the use of $l_1$
regularizers for the inverse Ising problem, which forces a
fraction of parameters to assume value zero, thus effectively reducing
the number of parameters. This approach is not appropriate here since
we are concerned with the accuracy (resp. divergence) of the strongest predicted
couplings; for our purposes
it makes no substantial difference whether weak couplings are inferred to
be small or set precisely to 0.  Our choice for $R$ is therefore
the simpler $l_2$ norm
\begin{equation}
\label{Rl2}
R_{l_2}(\boldsymbol{\mathrm{h}},\boldsymbol{\mathrm{J}})
=\lambda_h \sum\limits^N_{r=1} ||\boldsymbol{\mathrm{h}}_r||^2_2 
+ \lambda_J \sum_{1\leq i<j\leq N} 
||\boldsymbol{\mathrm{J}}_{ij}||^2_2,
\end{equation}
using two regularization parameters $\lambda_h$ and $\lambda_J$ for
field and coupling parameters.  An advantage of a regularizer is
that it eliminates the need to fix a gauge, since among all parameter
sets related by a gauge transformation, i.e., all parameter sets
resulting in the same Potts model, there will be exactly one set which
minimizes the strictly convex regularizer. For the case of
the $l_2$ norm, it can be shown that this leads to a gauge where
$\sum_{s=1}^q J_{ij}(k,s)=\frac{\lambda_h}{\lambda_J} h_i(k)$, $\sum_{s=1}^q J_{ij}(s,l)=\frac{\lambda_h}{\lambda_J}h_j(l)$, and $\sum_{s=1}^q h_i(s)=0$ for all $i$, $j (>i)$, $k$, and $l$.

To summarize this discussion: For NMFI, we regularize with
pseudocounts under the gauge constraints Eq.~(\ref{zero_gauge}). For
PLM, we regularize with $R_{l_2}$ under the full parametrization.

\subsection{Sequence reweighting}

Maximum-likelihood inference of Potts models relies -- as discussed above -- 
on the assumption that the $B$ sample configurations in our
data set are independently generated from
Eq.~(\ref{potts_model}). This assumption is not correct for biological
sequence data, which have a {\it phylogenetic} bias. In particular, 
in the databases there are many protein sequences from related species, 
which did not have enough
time of independent evolution to reach statistical
independence. Furthermore, the selection of sequenced species in the
genomic databases is dictated by human interest, and not by the aim to
have an as independent as possible sampling in the space of all
functional amino-acid sequences. A way to mitigate effects of uneven
sampling, employed in \cite{pmid22106262}, is to equip each sequence
$\boldsymbol{\sigma}^{(b)}$ with a \textit{weight} $w_b$ which
regulates its impact on the parameter estimates. Sequences deemed
unworthy of independent-sample status (too similar to other sequences)
can then have their weight lowered, whereas sequences which are quite
different from all other sequences will contribute with a higher
weight to the amino-acid statistics.

A simple but efficient way (cf. \cite{pmid22106262}) is to measure the
similarity $\hbox{sim}(\boldsymbol{\sigma}^{(a)},\boldsymbol{\sigma}^{(b)})$
of two sequences $\boldsymbol{\sigma}^{(a)}$ and
$\boldsymbol{\sigma}^{(b)}$ as the fraction of conserved positions
(i.e., identical amino acids), and compare this fraction to a
preselected threshold $x$ , $0<x<1$. The weight given to a sequence
$\boldsymbol{\sigma}^{(b)}$ can then be set to $w_b=\frac{1}{m_b}$, where $m_b$
is the number of sequences in the MSA similar to
$\boldsymbol{\sigma}^{(b)}$:
\begin{equation}
\label{inverse_weights}
m_b=|\{a \in \{1,...,B\} :
\hbox{sim}(\boldsymbol{\sigma}^{(a)},\boldsymbol{\sigma}^{(b)}) \geq x\}|.
\end{equation}
In~\cite{pmid22106262}, a suitable threshold $x$ was found to be $0.8$, results
only weakly dependent on this choice throughout $0.7<x<0.9$. We have
here followed the same procedure using threshold $x=0.9$.  
The corresponding reweighted frequency counts then become
\begin{eqnarray}
\label{pseudo_rew_frequencies}
f_i(k)&=&
\frac{1}{\lambda+B_{eff}} \left[ \frac{\lambda}{q}+\sum^B_{b=1} w_b \,
\delta(\sigma^{(b)}_i,k) \right] , \\
f_{ij}(k,l)&=&
\frac{1}{\lambda+B_{eff}} \left[ \frac{\lambda}{q^2}+\sum^B_{b=1} w_b \, 
\delta(\sigma^{(b)}_i,k)\, \delta(\sigma^{(b)}_j,l) \right] ,
\nonumber
\end{eqnarray}
where $B_{eff}=\sum^B_{b=1} w_b$ becomes a measure of the number of
effectively nonredundant sequences.

In the pseudolikelihood we use the direct analog of
Eq.~(\ref{pseudo_rew_frequencies}), i.e.,
\begin{eqnarray}
\label{reweighted_npll}
&&l_{pseudo}(\boldsymbol{\mathrm{h}},\boldsymbol{\mathrm{J}}) \\ 
&=&\!-\frac{1}{B_{eff}} \! \sum\limits^B_{b=1} \! w_b \! \sum\limits^N_{r=1} 
\log \left[ P_{\{\boldsymbol{\mathrm{h}}_r,\boldsymbol{\mathrm{J}}_r \}}(
\sigma_r=\sigma^{(b)}_r|\boldsymbol{\sigma}_{\backslash r}
=\boldsymbol{\sigma}^{(b)}_{\backslash r}) \right].
\nonumber
\end{eqnarray}
As in the frequency counts, each sequence is considered to contribute
a weight $w_b$, instead of the standard weight one used in
i.i.d. samples.

\subsection{Interaction scores}
In the inverse Ising problem, each interaction is scored by one scalar
coupling strength $J_{ij}$. These can easily be ordered, e.g. by
absolute size. In the inverse Potts problem, each position pair
$(i,j)$ is characterized by a whole $q\times q$ matrix
$\boldsymbol{\mathrm{J}}_{ij}$, and some scalar score $\mathcal{S}_{ij}$ is needed in
order to evaluate the `coupling strength' of two sites.

In \cite{Weigt} and \cite{pmid22106262} the score used is based on {\it
  direct information} (DI), i.e., the mutual information of a
restricted probability model not including any indirect coupling
effects between the two positions to be scored. Its construction
goes as follows: For each position pair $(i,j)$, (the estimate of)
$\boldsymbol{\mathrm{J}}_{ij}$ is used to set up a 'direct
distribution' involving only nodes $i$ and $j$,
\begin{equation}
\label{direct_prob}
P^{(dir)}_{ij}(k,l) \sim \exp 
\left ( J_{ij}(k,l)+h'_{i,k}+h'_{j,l} \right ).
\end{equation}
$h'_{i,k}$ and $h'_{j,l}$ are new fields, computed as to ensure
agreement of the marginal single-site distributions with the empirical
individual frequency counts $f_i(k)$ and $f_j(l)$. The score $\mathcal{S}_{ij}^{DI}$ is now
calculated as the mutual information of $P^{(dir)}$:
\begin{equation}
\label{direct_information}
\mathcal{S}_{ij}^{DI}=\sum\limits^{q}_{k,l=1} P^{(dir)}_{ij}(k,l) \log 
\left ( \frac{P^{(dir)}_{ij}(k,l)}{ f_i(k)\,f_j(l)}\right ).
\end{equation}
A nice characteristic of $\mathcal{S}_{ij}^{DI}$ is its invariance with respect to the
gauge freedom of the Potts model, i.e., both choices
Eqs.~(\ref{sum_gauge}) and (\ref{zero_gauge}) (or any other valid
choice) generate identical $\mathcal{S}_{ij}^{DI}$.

In the pseudolikelihood approach, we prefer not to use $\mathcal{S}_{ij}^{DI}$, as this
would require a pseudocount $\lambda$ to regularize the frequencies in
(\ref{direct_information}), introducing a third regularization variable in
addition to $\lambda_h$ and $\lambda_J$. Another possible scoring
function, already mentioned but not used in \cite{Weigt}, is the
\textit{Frobenius norm} (FN)
\begin{equation}
\label{frob_norm}
\| {\boldsymbol{\mathrm{J}}_{ij}}\|_2
=\sqrt{\sum\limits^{q}_{k,l=1} J_{ij}(k,l)^2}.
\end{equation}
Unlike $\mathcal{S}_{ij}^{DI}$, (\ref{frob_norm}) is \textit{not} independent of
gauge choice, so one must be a bit careful. 
As was noted in \cite{Weigt}, the zero sum gauge (\ref{sum_gauge}) minimizes
the Frobenius norm, in a sense making (\ref{sum_gauge}) the
most appropriate gauge choice for the score (\ref{frob_norm}).
Recall from above that our pseudolikelihood uses the full representation and 
fixes the gauge by the regularization terms $R_{l_2}$. 
Our procedure is therefore to first infer the interaction parameters
using the pseudolikelihood and the regularization, and
then to change to the zero-sum gauge:
\begin{equation}
\label{gauge_change}
J'_{ij}(k,l)= J_{ij}(k,l)-J_{ij}(\cdot,l)-J_{ij}(k,\cdot)+J_{ij}(\cdot,\cdot),
\end{equation}
where `$\cdot$' denotes average over the concerned position. One can
show that (\ref{gauge_change}) preserves the probabilities of
(\ref{potts_model}) (after altering the fields appropriately) and that
$J'_{ij}(k,l)$ satisfy (\ref{sum_gauge}).  A
possible Frobenius norm score is hence
\begin{equation}
\label{frob_norm2} 
\mathcal{S}_{ij}^{FN}=\| {\boldsymbol{\mathrm{J}}'_{ij}}\|_2=\sqrt{\sum\limits^{q}_{k,l=1} 
J'_{ij}(k,l)^2}.
\end{equation}
Lastly, we borrow an idea from Jones \textit{et
  al.}~\cite{journals/bioinformatics/JonesBCP12}, whose PSICOV method
also uses a norm rank ($1$-norm instead of Frobenius norm), but
adjusted by an \textit{average-product correction} (APC)
term. APC was introduced in \cite{journals/bioinformatics/DunnWG08} to
suppress effects from phylogenetic bias and insufficient
sampling. Incorporating also this correction, we have our scoring
function
\begin{equation}
\label{corr_norm}
\mathcal{S}_{ij}^{CN}=\mathcal{S}_{ij}^{FN}-\frac{\mathcal{S}_{\cdot j}^{FN}
\mathcal{S}_{i \cdot}^{FN}}{\mathcal{S}_{\cdot \cdot}^{FN}},
\end{equation}
where CN stands for `corrected norm'. An implementation of plmDCA in
MATLAB is available at \url{http://plmdca.csc.kth.se/}.

\section{Evaluating the performance of mfDCA and plmDCA across protein families}
\label{sec:exp}
We have performed numerical experiments using mfDCA and plmDCA on a
number of domain families from the Pfam database; here we report and
discuss the results.

\subsection{Domain families, native structures, and true-positive rates}
The speed of mfDCA enabled Morcos {\it et al.} \cite{pmid22106262} to
conduct a large-scale analysis using 131 families. PLM is
computationally more demanding than NMFI, so we chose to start with a
smaller collection of 17 families, listed in Table
\ref{tab:families}. To ease the numerical effort, we chose families
with relatively small $N$ and $B$. More precisely, we selected families
out of the first 115 Pfam entries (low Pfam ID), which have {\em (i)}
at most $N=130$ residues, {\em (ii)} between 2,000 and 22,000 sequences,
and {\em (iii)} reliable structural information (cf.~the PDB entries
provided in the table). No selection based on DCA performance was
done. In the appendix, a number of longer proteins is studied. The 
mfDCA performance on the selected families was found to be coherent 
with the larger data set of Morcos {\em et al.}.

\begin{table*}[htb]
\begin{center}
\begin{tabular}{|c|c|c|c|c|c|c|}
\hline
Family ID & $N$ & $B$ & $B_{eff}$ (90\%) & PDB ID& UniProt entry & UniProt residues \\ \hline
PF00011 & 102 & 5024 & 3481 & 2bol&TSP36\_TAESA&106-206 \\  \hline
PF00013 & 58 & 6059 & 3785 & 1wvn&PCBP1\_HUMAN&281-343\\ \hline
PF00014 & 53 & 2393 & 1812 & 5pti& BPT1\_BOVIN&39-91 \\ \hline
PF00017 & 77 & 2732 & 1741 & 1o47 &SRC\_HUMAN&151-233 \\ \hline
PF00018 & 48 & 5073 & 3354 & 2hda &YES\_HUMAN&97-144\\ \hline
PF00027 & 91 & 12129 & 9036 & 3fhi&KAP0\_BOVIN&154-238\\ \hline
PF00028 & 93 & 12628 & 8317 & 2o72 &CADH1\_HUMAN&267-366 \\ \hline
PF00035 & 67 & 3093 & 2254 & 1o0w &RNC\_THEMA&169-235 \\ \hline
PF00041 & 85 & 15551 & 10631 & 1bqu &IL6RB\_HUMAN&223-311\\ \hline
PF00043 & 95 & 6818 & 5141 & 6gsu &GSTM1\_RAT&104-192 \\ \hline
PF00046 & 57 & 7372 & 3314 & 2vi6 &NANOG\_MOUSE&97-153\\ \hline
PF00076 & 70 & 21125 & 14125 & 1g2e &ELAV4\_HUMAN&48-118 \\ \hline
PF00081 & 82 & 3229 & 1510 & 3bfr & SODM\_YEAST&27-115\\ \hline
PF00084 & 56 & 5831 & 4345 & 1elv &C1S\_HUMAN&359-421\\ \hline
PF00105 & 70 & 2549 & 1277 & 1gdc &GCR\_RAT&438-507\\ \hline
PF00107 & 130 & 17864 & 12114 & 1a71 &ADH1E\_HORSE&203-338\\ \hline
PF00111 & 78 & 7848 & 5805 & 1a70 &FER1\_SPIOL&58-132 \\
\hline
\end{tabular}
\end{center}
\caption{Domain families included in our study, listed with Pfam ID,
  length $N$, number of sequences $B$ (after removal of duplicate
  sequences), number of effective sequences $B_{eff}$ (under $x=0.9$,
  i.e., $90\%$ threshold for reweighting), and the PDB and UniProt
  specifications for the structure used to access the DCA prediction
  quality.}
\label{tab:families}
\end{table*}

\begin{figure*}[htb]
  \begin{center}
    \includegraphics[width=14cm,height=5cm]{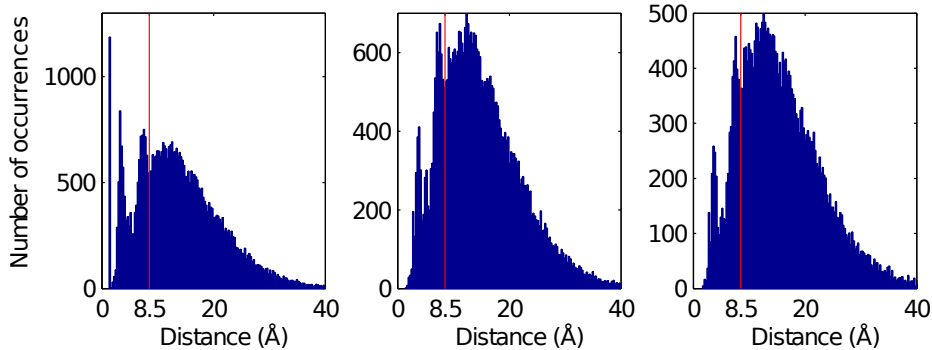}
  \end{center}
\caption{(Color online) Histograms of crystal-structure distances pooled from all
17 families, when including all pairs (left), pairs with $|j-i|>4$ (center), and pairs with $|j-i|>14$ (right). The vertical
line is our contact cutoff 8.5\AA.}
    \label{fig:0dist_hist}
\end{figure*}

To reliably assess how good a contact prediction is, something to
regard as a 'gold standard' is helpful. For each of the 17 families we
have therefore selected one representative high-resolution X-ray
crystal structure (resolution below 3\AA ), see 
Table \ref{tab:families} for the corresponding PDB identification. 

From these native protein structures, we have extracted
position-position distances $d(i,j)$ for each pair of sequence
positions, by measuring the minimal distance between any two heavy
atoms belonging to the amino acids present in these positions.
The left panel of Fig.~\ref{fig:0dist_hist} shows the distribution of these distances
in all considered families. Three peaks protrude from the background
distribution: one at small distances below 1.5\AA, a second at about
3-5\AA\ and a third at about 7-8\AA. The first peak corresponds to the
peptide bonds between sequence neighbors, whereas the other two peaks
correspond to nontrivial contacts between amino acids, which may be
distant along the protein backbone, as can be seen from the center and right panels
of Fig.~\ref{fig:0dist_hist}, which collect only distances between
positions $i$ and $j$ with minimal separation $|j-i|\geq 5$ resp.
$|j-i|\geq 15$. Following \cite{pmid22106262}, we take the peak at
3-5\AA\ to presumably correspond to short-range interactions like
hydrogen bonds or secondary-structure contacts, whereas the last peak
likely corresponds to long-range, possibly water-mediated
interactions. These peaks contain the nontrivial information we would
like to extract from sequence data using DCA. In order to accept the
full second peak, we have chosen a distance cutoff of 8.5\AA\ for true
contacts, slightly larger than the value of 8\AA\ used in
\cite{pmid22106262}.

Accuracy results are here reported primarily using
\textit{true-positive} (TP) rates, also the principal measurement in
\cite{pmid22106262} and
\cite{journals/bioinformatics/JonesBCP12}. The TP rate for $p$ is the
fraction of the $p$ strongest-scored pairs which are actually contacts
in the crystal structure, defined as described above. To exemplify TP rates, 
let us jump ahead and look at
Fig.~\ref{fig:1TP_rates}. For PLM and protein family PF00076, the TP
rate is 1 up to $p=80$, which means that all 80 top-$\mathcal{S}_{ij}^{CN}$ pairs are
genuine contacts in the crystal structure. At $p=200$, the TP rate has
dropped to 0.78, so $0.78 \cdot 200 = 156$ of the top 200 top-$\mathcal{S}_{ij}^{CN}$
pairs are contacts, while 44 are not.

\subsection{Parameter settings}
To set the stage for comparison, we started by running initial trials
on the 17 families using both NMFI and PLM with many different
regularization and reweighting strengths. Reweighting indeed raised
the TP rates, and, as reported in \cite{pmid22106262} for 131
families, results seemed robust toward the exact choice of the limit
$x$ around $0.7 \leq x \leq 0.9$. We chose $x=0.9$ to use throughout
the study.

In what follows, NMFI results are reported using the same list of
pseudocounts as in Fig.~S11 in \cite{pmid22106262}: $\lambda=w \cdot
B_{eff}$ with $w=\{0.11$, $0.25$, $0.43$, $0.67$, $1.0$, $1.5$, $2.3$,
$4.0$, $9.0\}$. During our analysis we also ran intermediate values,
and we found this covering to be sufficiently dense.  We give outputs
from two versions of NMFI: NMFI-DI and NMFI-DI(true). The former uses
pseudocounts for all calculations, whereas the latter switches to true
frequencies when it gets to the evaluations of
$\mathcal{S}_{ij}^{DI}$. We append 'DI' to the NMFI name, since, later
on, we will also try the $\mathcal{S}_{ij}^{CN}$ score for NMFI (which
we will call NMFI-CN).

With $l_2$ regularization in the PLM algorithm, outcomes were robust
against the precise choice of $\lambda_h$; TP rates were almost
identical when $\lambda_h$ was changed between $0.001$ and $0.1$.  We
therefore chose $\lambda_h=0.01$ for all experiments. What mattered,
rather, was the coupling regularization parameters $\lambda_J$, for
which we did a systematic scan from $\lambda_J=0$ and up using
step-size 0.005.

So, to summarize, the results reported here are based on $x=0.9$,
cutoff 8.5\AA, $\lambda_h=0.01$, and $\lambda$ and
$\lambda_J$ drawn from collections of values as described above.

\subsection{Main comparison of mfDCA and plmDCA}

\begin{figure}[hbt]
  \begin{center}
    \includegraphics[width=9cm,height=11cm]{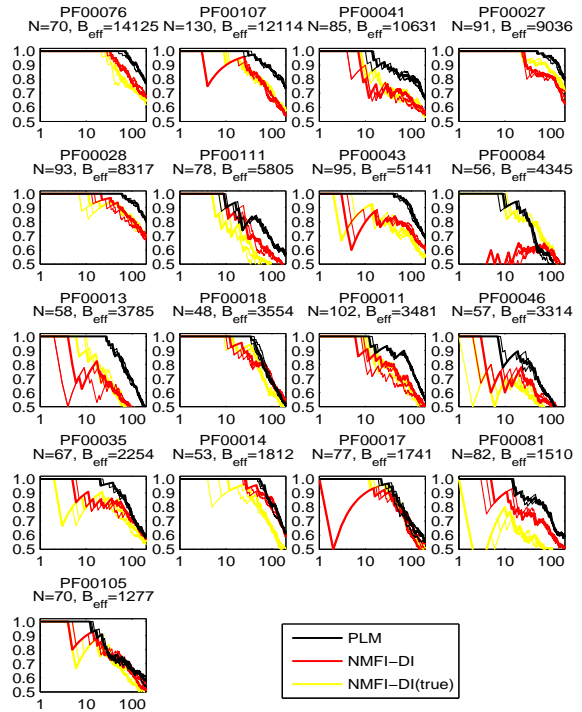}
  \end{center}
\caption{(Color online) Contact-detection results for the 17 families,
  sorted by $B_{eff}$. The y-axes are TP rates and x-axes are the
  number of predicted contacts $p$, based on pairs with $|j-i|>4$. The
  three curves for each method are the three regularization levels
  yielding highest TP rates across all families. The thickened curve
  highlights the best one out of these three ($\lambda=B_{eff}$ for
  NMFI and $\lambda_J=0.01$ for PLM).}
    \label{fig:1TP_rates}
\end{figure}

Fig.~\ref{fig:1TP_rates} shows TP rates for the different families and
methods. We see that the TP rates of plmDCA (PLM) are consistently
higher than those of mfDCA (NMFI), especially for families with
large $B_{eff}$. For what concerns the two NMFI versions:
NMFI-DI(true) avoids the strong failure seen in NMFI-DI for PF00084,
but for most other families, see in particular PF00014 and PF00081,
the performance instead drops using marginals without pseudocounts in
the $\mathcal{S}_{ij}^{DI}$ calculation. For both NMFI-DI and
NMFI-DI(true), the best regularization was found to be $\lambda=1
\cdot B_{eff}$, the same value as used in \cite{pmid22106262}. For
PLM, the best parameter choice was $\lambda_J=0.01$. Interestingly,
this same regularization parameter was optimal for basically all
families. This is somewhat surprising, since both $N$ and $B_{eff}$
span quite wide ranges ($48$-$130$ and $1277$-$14125$ respectively).

In the following discussion, we leave out all results for
NMFI-DI(true) and focus on PLM vs. NMFI-DI, i.e.,~the version used in
\cite{pmid22106262}. All plots remaining in this section use the
optimal regularization values: $\lambda=B_{eff}$ for NMFI and
$\lambda_J=0.01$ for PLM.

\begin{figure*}[hbt]
  \begin{center}
    \includegraphics[width=15cm,height=7cm]{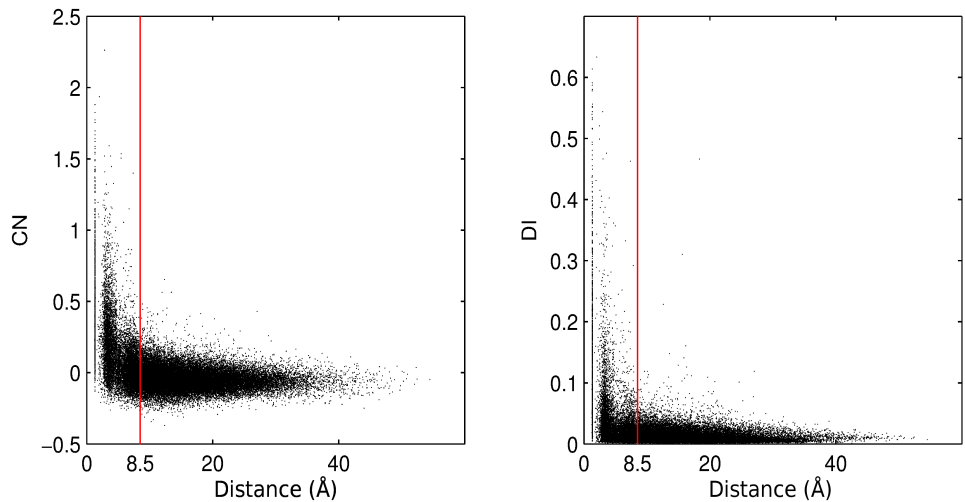}
  \end{center}
  \caption{(Color online) Score plotted against distance for all position pairs in
    all 17 families for PLM (left) and NMFI-DI (right). The vertical line is our contact cutoff at 8.5\AA.}
    \label{fig:1scores_vs_dist}
\end{figure*}

\begin{figure}[hbt]
  \begin{center}
    \includegraphics[width=9cm,height=11cm]{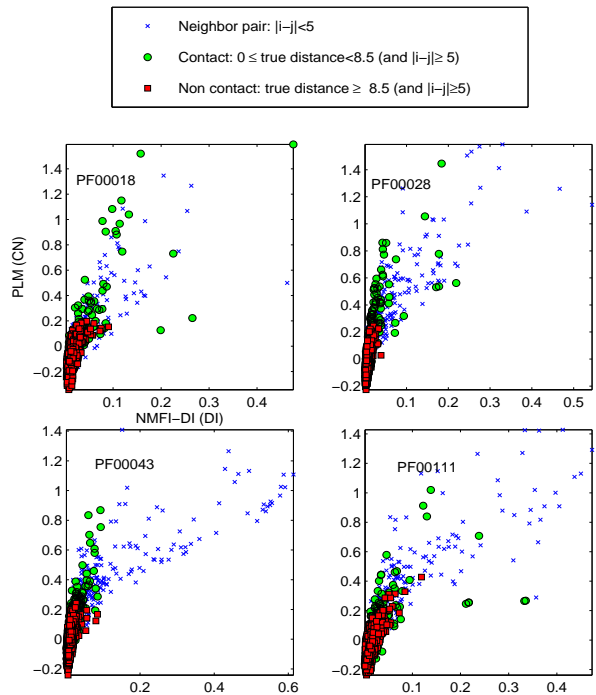}
  \end{center}
\caption{(Color online) Scatter plots of interaction scores for PLM and NMFI-DI from
  four families. For all plots, the axes are as indicated by the top
  left figure. The distance unit in the top box is \AA.}
    \label{fig:1scatter_plots}
\end{figure}

\begin{figure}[hbt]
  \begin{center}
    \includegraphics[width=9cm,height=11cm]{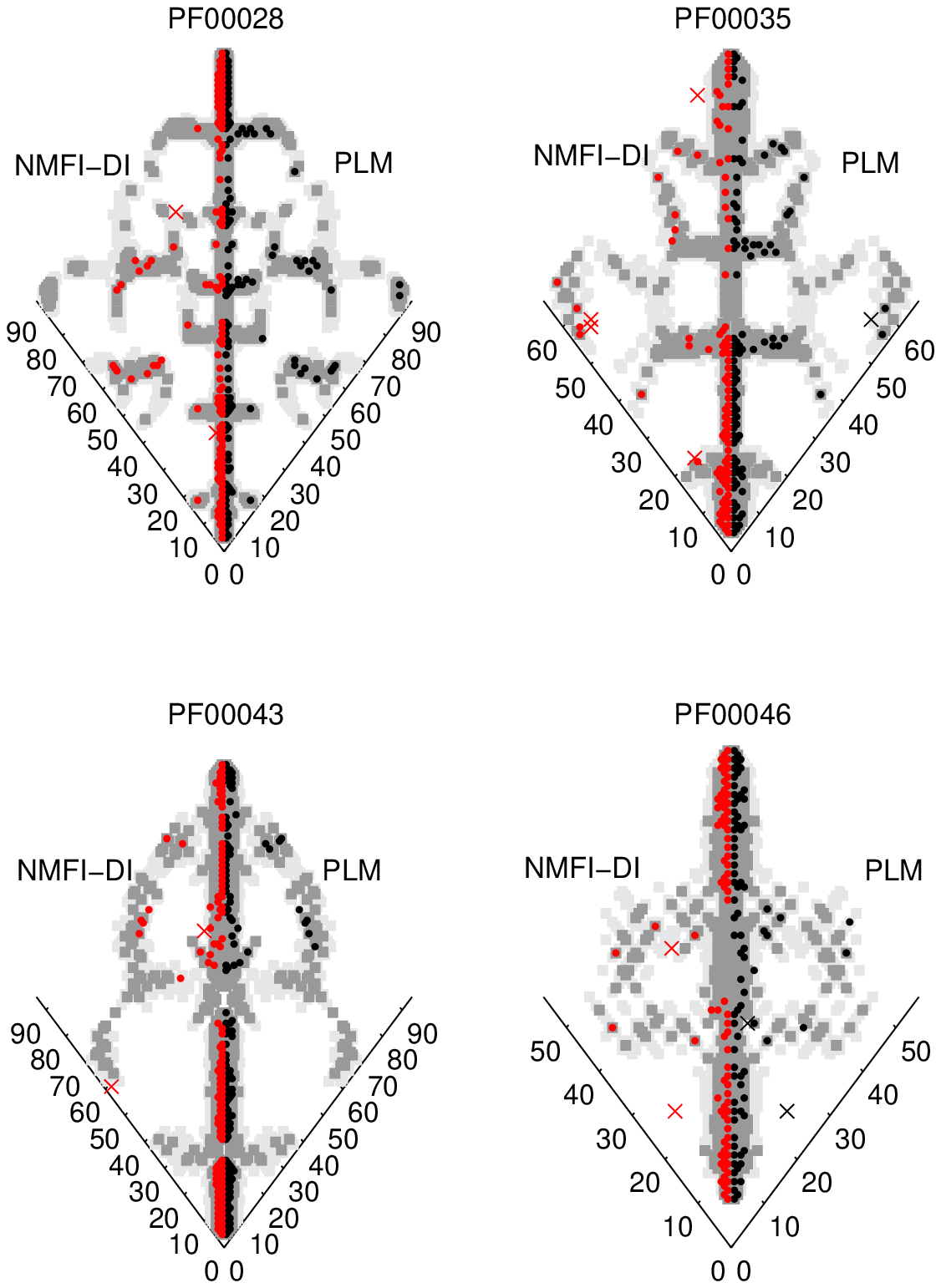}
  \end{center}
\caption{(Color online) Predicted contact maps for PLM (right part, black
  symbols) and NMFI-DI (left part, red symbols online) for four
  families. A pair $(i,j)$'s placement in the plots is found by
  matching positions $i$ and $j$ on the axes. Native contacts are
  indicated in gray. True and false positives are represented by
  circles and crosses, respectively. Each figure shows the $1.5N$
  strongest ranked pairs (including neighbors) for that family.}
    \label{fig:1contact_maps}
\end{figure}

TP rates only classify pairs as contacts ($d(i,j)<8.5$\AA) or
noncontacts ($d(i,j)\geq 8.5$\AA). To give a more detailed view of how
scores correlate with spatial separation, we show in
Fig.~\ref{fig:1scores_vs_dist} a scatter plot of the score
vs. distance for all pairs in all 17 families. PLM and NMFI-DI both
manage to detect the peaks seen in the true distance distribution of
Fig.~\ref{fig:0dist_hist}, in the sense that high scores are
observed almost exclusively at distances below 8.5\AA. Both methods
agree that interactions get, on average, progressively weaker going
from peak one, to two, to three, and finally to the bulk. 
We note that
the dots scatter differently across the PLM and NMFI-DI figures,
reflecting the two separate scoring techniques: $\mathcal{S}_{ij}^{DI}$ are strictly
nonnegative, whereas APC corrected norms can assume negative values.
We also observe how sparse the extracted signal is: most spatially
close pairs do not show elevated scores. However, from the other side,
almost all strongly coupled pairs are close, so the biological hypothesis of
Sec.~\ref{sec:psp} is well supported here.

Fig.~\ref{fig:1scatter_plots} shows scatter plots of scores for PLM
and NMFI-DI for some selected families. Qualitatively the same
patterns were observed for all families. The points are clearly
correlated, so, to some extent, PLM and NMFI-DI agree on the
interaction strengths. Due to the different scoring
schemes, we would not expect numerical coincidence of scores. Many
of PLM's top-scoring position pairs have also top scores for NMFI-DI
and vice versa. The largest discrepancy is in how much more strongly
NMFI-DI responds to pairs with small $|j-i|$; the crosses tend to
shoot out to the right. PLM agrees that many of these neighbor pairs
interact strongly, but, unlike NMFI-DI, it also shows rivaling
strengths for many $|j-i|>4$-pairs.


An even more detailed picture is given by considering {\it contact
  maps}, see Fig.~\ref{fig:1contact_maps}. The tendency observed in
the last scatter plots remains: NMFI-DI has a larger portion of highly
scored pairs in the neighbor zone, which are the middle stretches in these
figures. An important observation is, however, that clusters of
contacting pairs with long 1D sequence separation are captured by both
algorithms.

Note, that only a relatively small fraction of contacts is uncovered
by DCA, before false positives start to appear, and that many native
contacts are missed. However, the aim of DCA cannot be to reproduce a
complete contact map: It is based on sequence correlations alone ({\it
  e.g.} it cannot predict contacts for 100\% conserved residues), it
does not consider any physico-chemical property of amino acids (they
are seen as abstract letters), it does not consider the need to be
embeddable in 3D. Furthermore, proteins may undergo conformational
changes (e.g. allostery) or homo-dimerize, so coevolution may be
induced by pairs which are not in contact in the X-ray crystal
structure used for evaluating prediction accuracy. The important point
-- and this seems to be a distinguishing feature of maximum-entropy
models as compared to simpler correlation-based methods -- is to find
a sufficient number of well-distributed contacts to enable de-novo 3D
structure prediction
\cite{journals/corr/abs-1110-5091,Sulkowska2012,Marks2012-Cell,Schug2009,Dago2012,Nugent}.
In this sense, it is more important to achieve high accuracy of the
first predicted contacts, than intermediate accuracy for many
predictions.

In summary, the results suggest that the PLM method offers some
interesting progress compared to NMFI.  However, let us also note that
in the comparison we also had to change both scoring and
regularization styles. It is thus conceivable that a NMFI with the new
scoring and regularization could be more competitive with PLM. Indeed,
upon further investigation, detailed in Appendix \ref{app:extra2}, we
found that part of the improvement in fact does stem from the new
score. In Appendix \ref{app:extra3}, where we extend our results to
longer Pfam domains, we therefore add results from NMFI-CN, an updated
version of the code used in \cite{pmid22106262} which scores by
$\mathcal{S}_{ij}^{CN}$ instead of $\mathcal{S}_{ij}^{DI}$.

\subsection{Run times}
In general, NMFI, which is merely a matrix inversion, is very quick
compared with PLM; most families
in this study took only seconds to run through the NMFI code.

In contrast to the message-passing based method used in~\cite{Weigt},
a DCA using PLM is nevertheless feasible for all protein families in Pfam. 
The objective function in PLM is a sum over nodes and samples
and its execution time is therefore expected
to depend both on $B$ (number of members of
a protein family in Pfam) and $N$ (length of the aligned
sequences in a protein family). 

On a standard desktop computer, using a basic MATLAB-interfaced
C-implementation of conjugate gradient descent, the run times for
PF00014, PF00017, and PF00018 (small $N$ and $B$) were 50, 160 and 90  respectively. For PF00041 (small $N$ but larger $B$) one run
took 15 min.
For domains with larger $N$, like those in Appendix \ref{app:extra3}, run times grow approximately
apace. For example, the run times for PF00026 ($N=314$) and PF00006 ($N=215$)
were 80 and 65 min respectively.  

A well-known alternative use of pseudolikelihoods is to minimize each $g_r$ separately.
While slightly more crude, such an 'asymmetric' variant of plmDCA would amount to $N$ independent (multiclass) 
logistic regression problems, which would make parallel execution trivial on up to $N$ cores.
A rigorous examination of the asymmetric version's performance is beyond the scope of the present work, 
but initial tests suggest that even on a single processor it requires convergence times almost an order of 
magnitude smaller than the symmetric one (which we used in this work), while still yielding almost exactly the same TP rates. 
Using $N$ processors, the above mentioned run times could thus, in principle, 
be dropped by a factor as large as $10N$, suggesting that plmDCA can be made competitive not only in terms of accuracy, but also in terms of computational speed.

Finally, all of these times were obtained
cold-starting with all fields and couplings at 0.
Presumably, one can improve by using an appropriate initial guess
obtained, say, from NMFI. This has however not been implemented here.

\section{Discussion}
\label{sec:discussion}

In this work, we have shown that a direct-couping analysis built on
pseudolikelihood maximization (plmDCA) consistently outperforms the
previously described mean-field based analysis (mfDCA), as assessed
across a number of large protein-domain families. The
advantage of the pseudolikelihood approach was found to be partially
intrinsic, and partly contingent on using a sampling-corrected
Frobenius norm to score inferred direct statistical coupling matrices.

On one hand, this improvement might not be surprising: it is known
that for very large data sets PLM becomes
asymptotically equivalent to full maximum-likelihood inference,
whereas mean-field inference remains intrinsically approximate,
and this may result in an improved PLM performance also for finite data
sets \cite{1107.3536v2}.

On the other hand, the above advantage holds if and only
if the following two conditions are fulfilled: 
Data are drawn independently from a probability distribution,
and this probability distribution is the Boltzmann distribution of a Potts model.
None of
these two conditions actually hold for real protein
sequences. On artificial data, refined mean-field methods
(Thouless-Anderson-Palmer equations, Bethe approximation) also lead to
improved model inference as compared to NMFI,
cf.~{\it
  e.g.}~\cite{Frontiers,SessakMonasson,MezardMora,Nguyen-Berg2012a},
but no such improvement has been observed in real protein data
\cite{pmid22106262}. The results of the paper are therefore
interesting and highly nontrivial. They also suggest that other 
model-learning methods from statistics such as 'Contrastive Divergence'~\cite{hinton2002} or
the more recent 'Noise-Contrastive Estimation'~\cite{Gutmann2012a}, 
could be explored to further increase our capacity to extract structural information from
protein sequence data.

Disregarding the improvements, we find that 
overall the predicted contact pairs
for plmDCA and mfDCA are highly overlapping, illustrating the
robustness of DCA results with respect to the algorithmic
implementation. This observations suggests that, in the context of
modeling the sequence statistics by pairwise Potts models, most
extractable information might already be extracted from the MSA. However, it
may well also be that there is alternative information hidden in the
sequences, for which we would need to go beyond pairwise models, or integrate the
physico-chemical properties of different amino acids into the procedure, or extract
even more information from large sets of evolutionarily related
amino-acid sequences. DCA is only a step in this direction.

In our work we have seen that simple sampling corrections, more
precisely sequence reweighting and the average-product correction of
interaction scores, lead to an increased accuracy in predicting 3D
contacts of amino acids, which are distant on the protein's
backbone. It is, however, clear that these somewhat heuristic
statistical fixes cannot correct for the complicated hierarchical
phylogenetic relationships between proteins, and that more
sophisticated methods would be needed to disentangle phylogenetic from
functional correlations in massive sequence data. 
To do so is an open
challenge, which would leave the field of {\it equilibrium} inverse
statistical mechanics, but where methods of inverse statistical
mechanics may still play a useful role.
 

\section*{Acknowledgments}
This work was supported by the Academy of Finland as part of its
Finland Distinguished Professor program, project 129024/Aurell, and
through the Center of Excellence COIN. Discussions with S.~Cocco and
R.~Monasson are gratefully acknowledged.

\bibliography{refReport}

\appendix
\section{Circle plots}
\label{app:extra1}

\begin{figure}[hbt]
  \begin{center}
    \includegraphics[width=9cm,height=12cm]{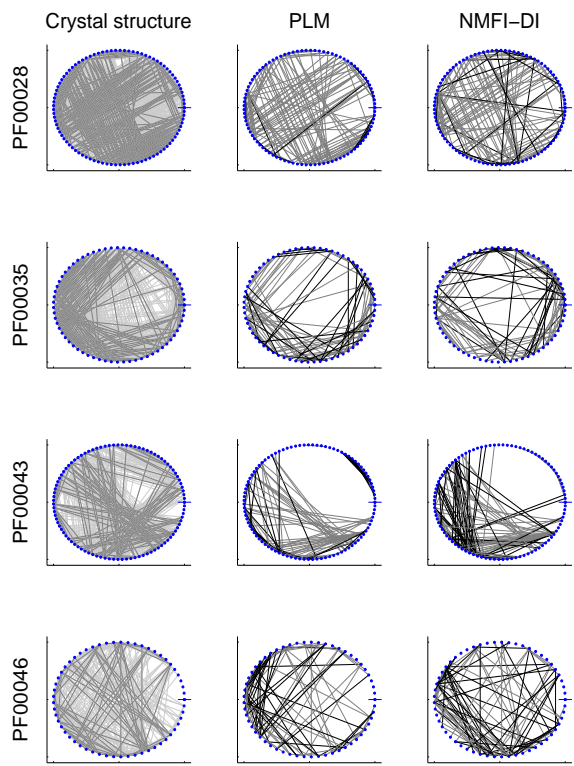}
  \end{center}
\caption{(Color online) Connections for four families overlaid on circles. Position
`1' is indicated by a dash. The leftmost column shows contacts in
the crystal structure (dark gray for $d(i,j)<5$\AA\ and light gray
for 5\AA$\leq d(i,j)<8.5$\AA). The other two columns show the top
$1.5N$ strongest ranked $|j-i|>4$-pairs for PLM and NMFI-DI, with
gray for true positives and black for false positives.}
    \label{fig:1circle_plots}
\end{figure}

To get a sense of how false positives distribute across the domains,
we draw interactions into circles in Fig.~\ref{fig:1circle_plots}.
Among erroneously predicted contacts there is some tendency towards
loopiness, especially for NMFI-DI; the black lines tend to `bounce
around' in the circles. It hence seems that relatively few nodes are
responsible for many of the false positives. We performed an
explicit check of the data columns belonging to these `bad' nodes,
and we found that they often contained strongly biased data, i.e.,
had a few large $\boldsymbol{\mathrm{f}}_i(k)$. In such cases, it
seemed that NMFI-DI was more prone than PLM to report a (predicted) interaction.

\section{Other scores for naive mean-field inversion}
\label{app:extra2}

\begin{figure}[hbt]
  \begin{center}
    \includegraphics[width=9cm,height=7cm]{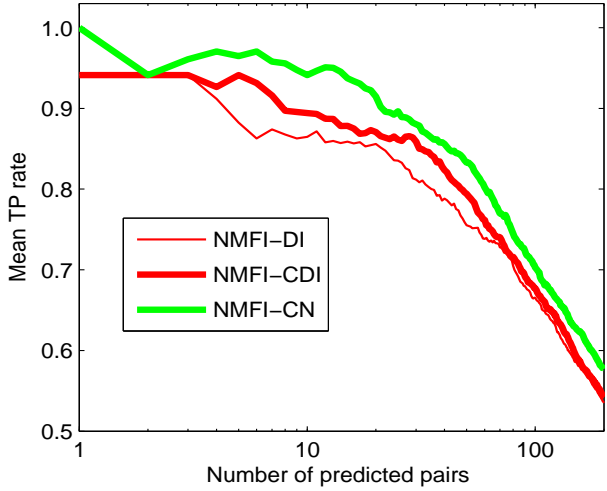}
  \end{center}
\caption{(Color online) Mean TP rates, using pairs with $|j-i|>4$, for NMFI with
old score $\mathcal{S}_{ij}^{DI}$, new APC score $\mathcal{S}_{ij}^{CDI}$, and
the norm score $\mathcal{S}_{ij}^{CN}$.  Each curve corresponds to the best $\lambda$ for
that particular score.}
    \label{fig:2mean_TP_rates}
\end{figure}

We also investigated NMFI performance using the APC term for the $\mathcal{S}_{ij}^{DI}$
scoring and using our new $\mathcal{S}_{ij}^{CN}$ score. In the second case we first
switch the parameter constraints from (\ref{zero_gauge}) to
(\ref{sum_gauge}) using (\ref{gauge_change}). Mean TP rates using
the modified score are shown in Fig. \ref{fig:2mean_TP_rates}.
We observe that APC in $\mathcal{S}_{ij}^{DI}$ scoring increases TP rates slightly, while
$\mathcal{S}_{ij}^{CN}$ scoring can improve TP rates overall. We remark, however, that for the second-highest ranked interaction ($p=2$) NMFI with the
original $\mathcal{S}_{ij}^{DI}$ (NMFI-DI) ties with NMFI with $\mathcal{S}_{ij}^{CN}$
(NMFI-CN).

\begin{figure*}[hbt]
  \begin{center}
    \includegraphics[width=15cm,height=7cm]{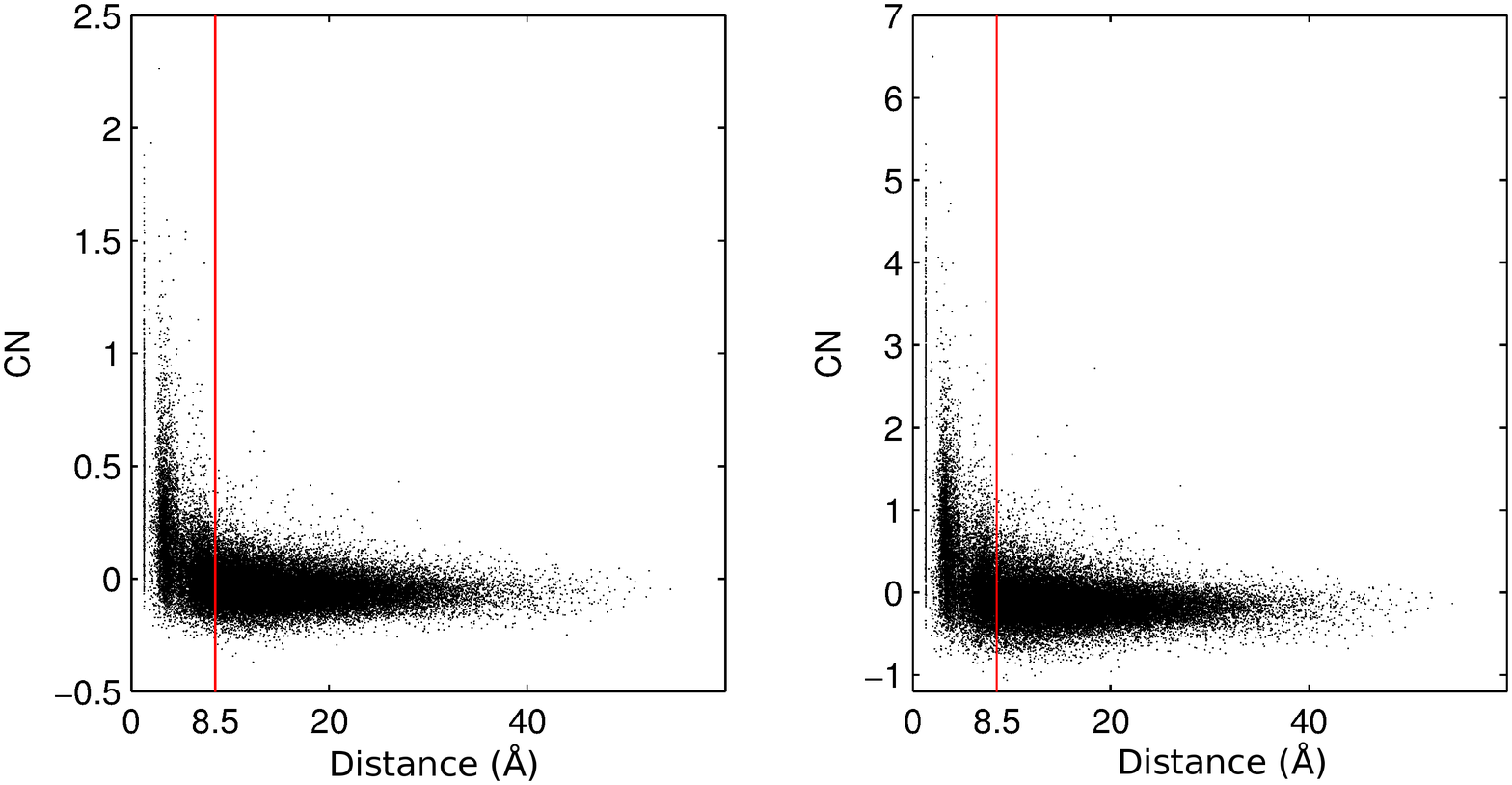}
  \end{center}
  \caption{(Color online) Score plotted against distance for all position pairs in
    all 17 families for PLM (left) and NMFI-CN (right). The vertical line is our contact cutoff at 8.5\AA.}
    \label{fig:3scores_vs_dists}
\end{figure*}

Motivated by the results of Fig. \ref{fig:2mean_TP_rates}, we
decided to compare NMFI and PLM under the $\mathcal{S}_{ij}^{CN}$ score. All figures in
this paragraph show the best regularization for each method, unless
otherwise stated. Figure \ref{fig:3scores_vs_dists} shows score vs.
distance for all pairs in all families. Unlike Fig.
\ref{fig:1scores_vs_dist}, the two plots now show very similar
profiles. We note, however, that NMFI's $\mathcal{S}_{ij}^{CN}$ scores trend two
to three times larger than PLM's (the scales on the vertical axes
are different). Perhaps this is an inherent feature of these methods, or
simply a consequence of the different types of regularization.

\begin{figure}[H]
  \begin{center}
    \includegraphics[width=9cm,height=11cm]{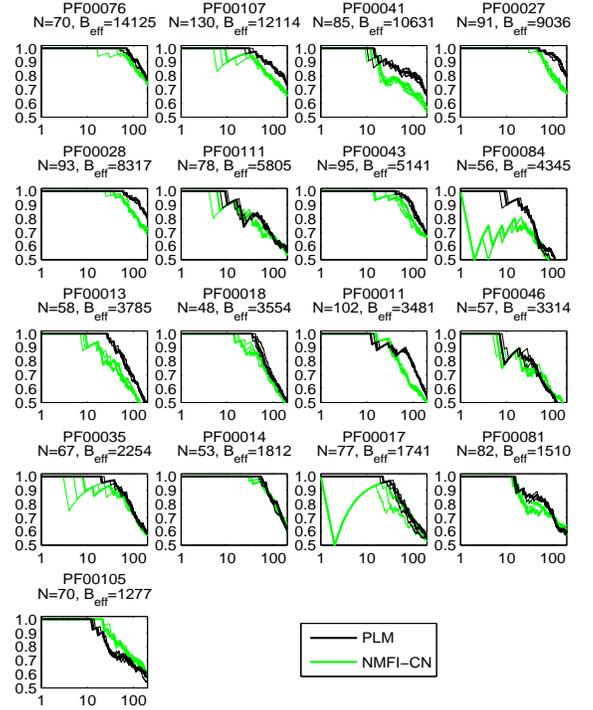}
  \end{center}
\caption{(Color online) Contact-detection results for all the families in our study
(sorted by $B_{eff}$), now with the $\mathcal{S}_{ij}^{CN}$ score for NMFI. The y-axes are TP
rates and the x-axes are the number of predicted contacts $p$, based on
pairs with $|j-i|>4$. The three curves for each method are the three
regularization levels yielding highest TP rates across all families.
The thickened curve highlights the best one out of these three
($\lambda=2.3 \cdot B_{eff}$ for NMFI-CN and $\lambda_J=0.01$ for
PLM).}
    \label{fig:3TP_rates}
\end{figure}

Figure~\ref{fig:3TP_rates} shows the
same situation as Fig.~\ref{fig:1TP_rates}, but using $\mathcal{S}_{ij}^{CN}$ to score NMFI. 
The three best
regularization choices for NMFI-CN turned out the same as before,
i.e., $\lambda=1 \cdot B_{eff}$, $\lambda=1.5 \cdot B_{eff}$ and
$\lambda=2.3 \cdot B_{eff}$, but the best out of these three was now
$\lambda=2.3 \cdot B_{eff}$ (instead of $\lambda=1 \cdot B_{eff}$).
Comparing  Fig.~\ref{fig:1TP_rates} and  Fig.~\ref{fig:3TP_rates} 
one can see that the difference between the two methods is now
smaller; for several families, the
prediction quality is in fact about the same for both methods. Still,
PLM maintains a somewhat higher TP rates overall.

\begin{figure}[H]
  \begin{center}
    \includegraphics[width=9cm,height=11cm]{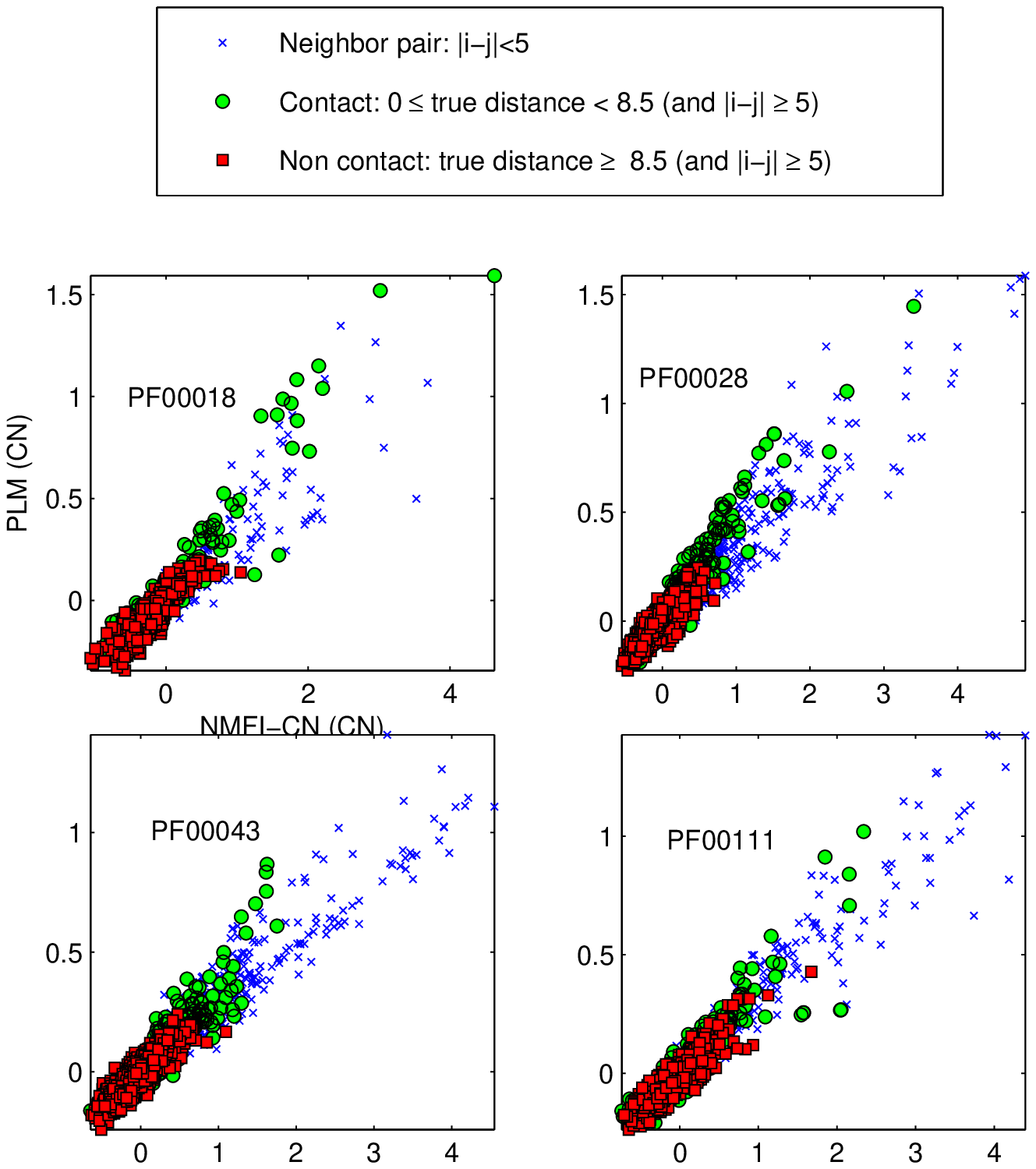}
  \end{center}
\caption{(Color online) Scatter plots of interaction scores for PLM and NMFI-CN
from four families. For all plots, the axes are as indicated by the
top left figure. The distance unit in the top box is \AA.}
    \label{fig:3scatter_plots}
\end{figure}

Figure \ref{fig:3scatter_plots} shows
scatter plots for the same families as in Fig.~\ref{fig:1scatter_plots}
but using the $\mathcal{S}_{ij}^{CN}$ scoring for NMFI. The points
now lie more clearly on a line, from which we conclude that
the bends in Fig.~\ref{fig:1scatter_plots} were likely 
a consequence of differing scores. Yet, the trends seen in 
Fig. \ref{fig:1scatter_plots} remain:
NMFI gives more attention to neighbor pairs than does PLM.

\begin{figure}[H]
  \begin{center}
    \includegraphics[width=9cm,height=11cm]{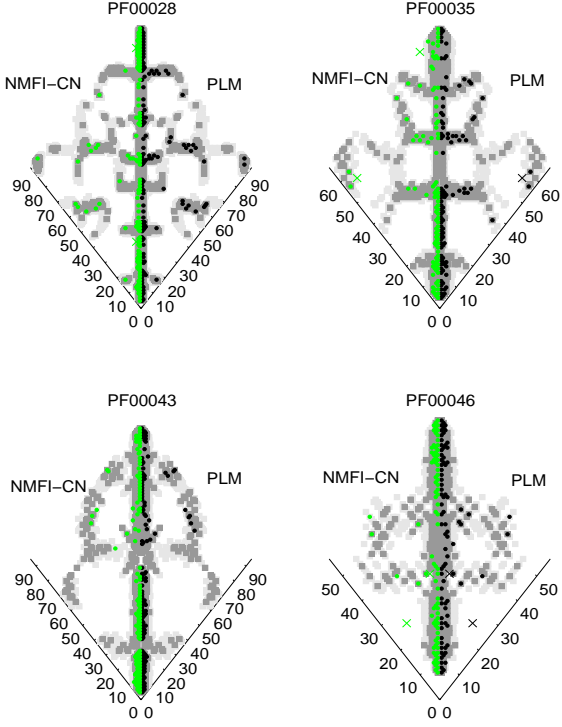}
  \end{center}
\caption{(Color online) Predicted contact maps for PLM (right part, black symbols) and NMFI-CN (left part, green symbols online) from four families. A pair
$(i,j)$'s placement in the plots is found by matching positions $i$
and $j$ on the axes. Contacts are indicated by gray (dark for
$d(i,j)<5$\AA\ and light for 5\AA$\leq d(i,j)<8.5$\AA). True and
false positives are represented by circles and crosses,
respectively. Each figure shows the $1.5N$ strongest ranked pairs
(including neighbors) for that family.}
    \label{fig:3contact_maps}
\end{figure}

In Fig. \ref{fig:3contact_maps}, we
recreate the contact maps of Fig. \ref{fig:1contact_maps} with
NMFI-CN in place of NMFI-DI and find that the plots are more symmetric.
As expected, asymmetry is seen primarily for small $|j-i|$; NMFI tends
to crowd these regions with lots of loops.

To investigate why NMFI
assembles so many top-scored pairs in certain neighbor regions, we
performed an explicit check of the associated MSA columns. A
relevant regularity was observed: when gaps appear in a sequence,
they tend to do so in long strands. The picture 
can be illustrated by 
the following hypothetical MSA (in our implementation, the gap state is 1):
\[
 \begin{pmatrix}
\cdots \\
\cdots $ 6 $ 5 $ 9 $ 7 $ 2 $ 6 $ 8 $ 7 $ 4 $ 4 $ 2 $ 2 \cdots\\
\cdots $ 1 $ 1 $ 1 $ 1 $ 1 $ 1 $ 1 $ 1 $ 1 $ 1 $ 2 $ 8 \cdots\\
\cdots $ 6 $ 5 $ 2 $ 7 $ 2 $ 3 $ 8 $ 9 $ 5 $ 4 $ 2 $ 3 \cdots\\
\cdots $ 3 $ 7 $ 4 $ 7 $ 2 $ 6 $ 8 $ 7 $ 9 $ 4 $ 2 $ 3 \cdots\\
\cdots $ 3 $ 7 $ 4 $ 7 $ 2 $ 3 $ 8 $ 8 $ 9 $ 4 $ 2 $ 9 \cdots\\
\cdots $ 1 $ 1 $ 1 $ 1 $ 1 $ 1 $ 1 $ 4 $ 5 $ 4 $ 2 $ 9 \cdots\\
\cdots $ 8 $ 5 $ 9 $ 7 $ 2 $ 9 $ 8 $ 7 $ 4 $ 4 $ 2 $ 4 \cdots\\
\cdots $ 1 $ 1 $ 1 $ 1 $ 1 $ 1 $ 1 $ 1 $ 1 $ 1 $ 2 $ 4 \cdots\\
\cdots
 \end{pmatrix}.
\]
We recall that gaps ('1' states)
are necessary for satisfactory alignment of the sequences in a family 
and that in our procedure we treat gaps 
just another amino acid, with its associated interaction parameters.
We then make the obvious observation that independent samples
from a Potts model will only contain long subsequences of
the same state with low probability. In other words, 
the model to which we fit the data cannot describe long
stretches of '1' states, which is a feature of the data.
It is hence quite conceivable that the two methods handle
this discrepancy between data and models differently since
we do expect this gap effect to generate 
large $J_{ij}(1,1)$
for at least some pairs with small $|j-i|$. 


\begin{figure}[H]
  \begin{center}
    \includegraphics[width=9cm,height=13cm]{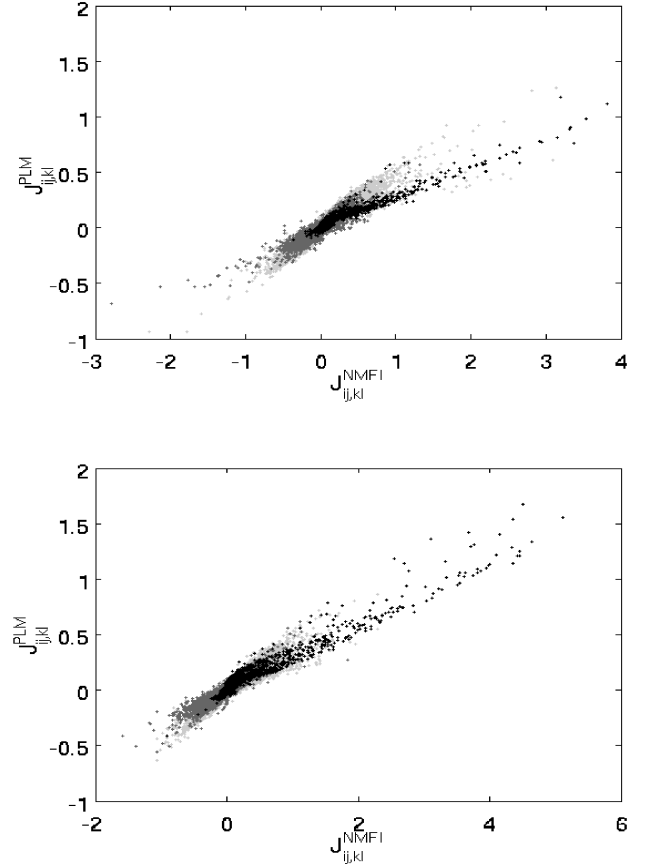}
  \end{center}
\caption{Scatter plots of estimated
$J_{ij,kl}=J_{ij}(k,l)$ from PF00014 (top) and
PF00043 (bottom). Black dots are `gap--gap' interactions ($k=l=1$), dark gray
dots are `gap--amino-acid' interactions ($k=1$ and $l\neq 1$, or
$k\neq 1$ and $l=1$), and light gray dots are `amino-acid--amino-acid'
interactions ($k\neq 1$ and $l\neq 1$).}
    \label{fig:4Jscatter}
\end{figure}

Figure \ref{fig:4Jscatter} shows scatter plots for all coupling
parameters $J_{ij}(k,l)$ in PF00014, which has
a modest amount of gap sections, and in PF00043, which has
relatively many. As outlines above, the
$J_{ij}(1,1)$-parameters are among the largest
in magnitude, especially for PF00043. We also note that the black dots steer to
the right; NMFI clearly reacts more strongly to the gap-gap interactions
than PLM.

Jones \textit{et al.} \cite{journals/bioinformatics/JonesBCP12} disregarded contributions from gaps in their scoring
by simply skipping the gap state when doing their norm summations.
We tried this but found no significant improvement for either
method. The change seemed to affect only pairs with small $|j-i|$
(which is reasonable), and our TP rates are based on pairs with
$|j-i|>4$. If gap interactions are indeed responsible for reduced
prediction qualities, removing their input during scoring is just a
Band-Aid type solution. A better way would be to suppress them
already in the parameter estimation step. That way, all interplay
would have to be accounted for without them. Whether or not 
there are ways to effectively handle the inference problem in PSP by
ignoring gaps or treating them differently, is an issue
which goes beyond the scope of this work.

\begin{figure}[H]
  \begin{center}
    \includegraphics[width=9cm,height=11cm]{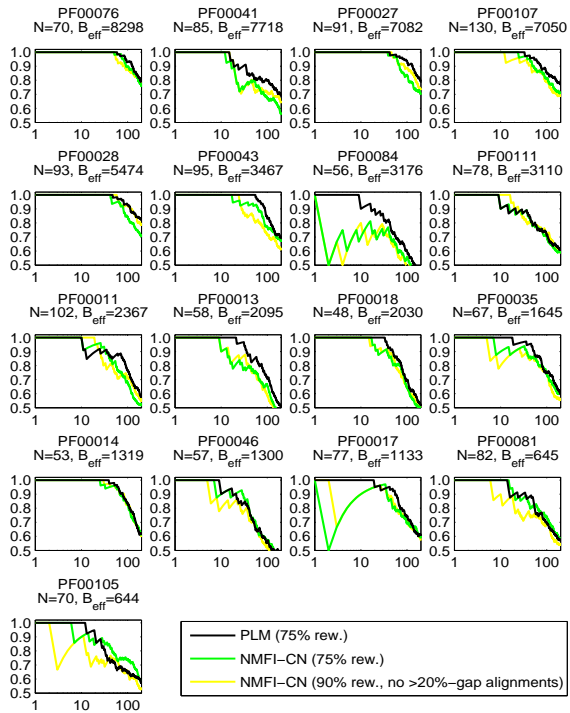}
  \end{center}
\caption{(Color online) Contact-detection results for all the families in our
study. The y-axes are TP rates and x-axes are the number of predicted
contacts $p$, based on pairs with $|j-i|>4$. Two curves are included for reweighting margin $x=0.75$, and one
for reweighting margin $x=0.9$ after deletion of all sequences with
more than $20\%$ gaps. Results for each method correspond to the
regularization level yielding highest TP rates across all families
($\lambda=2.3 \cdot B_{eff}$ for both NMFI-CN and $\lambda_J=0.01$
for PLM).}
    \label{fig:4TP_rates}
\end{figure}

We also investigated whether the gap effect depends on the 
sequence similarity reweighting factor $x$, which up to 
here was chosen $x=0.9$.
Perhaps the gap effect can be dampened by stricter definition of
sequence uniqueness? In Fig.~\ref{fig:4TP_rates} we show another
set of TP rates, but now for $x=0.75$. We also include results for NMFI run on
alignment files from which all sequences with more than $20\%$ gaps
have been removed. The best regularization choice for each method
turned out the same as in Fig. \ref{fig:3TP_rates}: $\lambda=2.3 \cdot
B_{eff}$ for NMFI-CN and $\lambda_J=0.01$ for PLM.  Overall, PLM maintains
the same advantage over NMFI-CN it had in
Fig.~\ref{fig:3TP_rates}. Removing gappy sequences seems to trim down
more TP rates than it raises, probably since useful information in the
nongappy parts is discarded unnecessarily.

\section{Extension to 28 protein families}
\label{app:extra3}
To sample a larger set of families, we conducted an additional
survey of 28 families, now covering lengths across the wider range of 50-400.
The list is given in Table~\ref{tab:morefamilies}. 
We here kept the reweighting level
at $x=0.8$ as in \cite{pmid22106262}, while the TP rates were again
calculated using the cutoff 8.5\AA. 
The
pseudocount strength for NMFI was varied in the same interval as in the main text.
We did not try to optimize the PLM regularization parameters for this trial,
but merely used $\lambda_h=\lambda_J=0.01$ as determined for the smaller families in the main text.

\begin{figure}[H]
  \begin{center}
    \includegraphics[width=9cm,height=6cm]{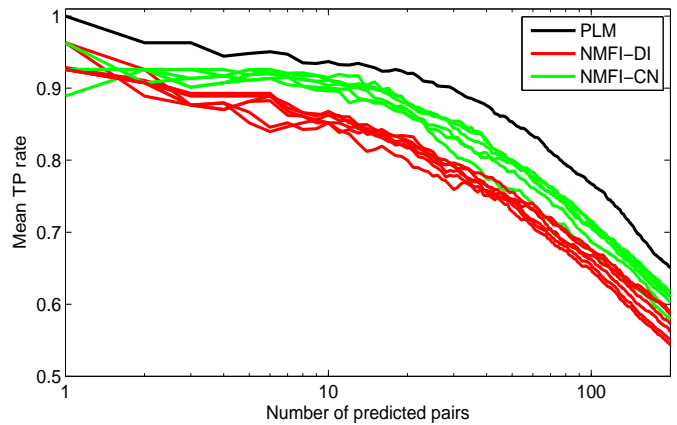}
  \end{center}
  \caption{(Color online) Mean TP rates over the larger set of 28 families for PLM with $\lambda_J=0.01$ and varying regularization values for NMFI-CN and NMFI-DI.}
    \label{fig:meanTPall}
\end{figure}

Figure~\ref{fig:meanTPall} 
shows qualitatively the same behavior as in the smaller set
of families: TP rates increase partly from changing 
from the $\mathcal{S}_{ij}^{DI}$ score to the $\mathcal{S}_{ij}^{CN}$ score, and partly from changing 
from NMFI to PLM.
Our positive results thus do not seem to
be particular to short-length families.

\begin{figure}[H]
  \begin{center}
    \includegraphics[width=9cm,height=6cm]{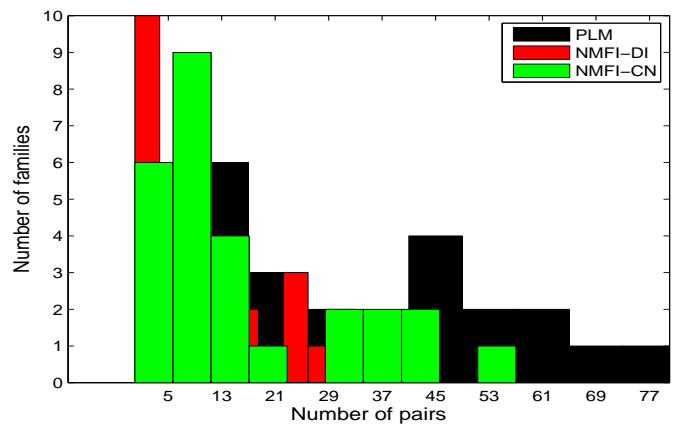}
  \end{center}
  \caption{(Color online) Distribution of 'perfect' accuracy for the three methods. The x-axis shows the number of top-ranked pairs for which the TP rates stays at one, and the y-axis shows the number of families.}
    \label{fig:distribution}
\end{figure}

Apart from the average TP rate for each value of $p$ ($p$'th strongest
predicted interactions) one can also evaluate performance by different 
criteria. In this larger survey we investigated the distribution of 
values of $p$ such that the TP rate in a family is one.
Fig.~\ref{fig:distribution} shows the histograms of the number of families
for which the top $p$ predictions are correct, clearly showing that
the difference between PLM and NMFI (using the two scores) primarily
occurs at the high end. The difference in average performance between
PLM and NMFI at least partially stems from PLM getting more strongest
contact predictions with 100\% accuracy.


\begin{table*}[hbt]
\begin{tabular}{|c|c|c|c|c|c|c|}
\hline
ID & N & B & $B_{eff}(80\%)$ & PDB ID & UniProt entry & UniProt residues\\
\hline
PF00006&215&10765&641&2r9v&ATPA\_THEMA&149-365\\
\hline
PF00011&102&5024&2725&2bol&TSP36\_TAESA&106-206\\
\hline
PF00014&53&2393&1478&5pti&BPT1\_BOVIN&39-91\\
\hline
PF00017&77&2732&1312&1o47&SRC\_HUMAN&151-233\\
\hline
PF00018&48&5073&335&2hda&YES\_HUMAN&97-144\\
\hline
PF00025&175&2946&996&1fzq&ARL3\_MOUSE&3-177\\
\hline
PF00026&314&3851&2075&3er5&CARP\_CRYPA&105-419\\			 
\hline
PF00027&91&12129&7631&3fhi&KAP0\_BOVIN&154-238\\
\hline			
PF00028&93&12628&6323&2o72&CADH1\_HUMAN&267-366\\
\hline	
PF00032&102&14994&684&1zrt&CYB\_RHOCA&282-404\\
\hline
PF00035&67&3093&1826&1o0w&RNC\_THEMA&169-235\\
\hline
PF00041&85&15551&8691&1bqu&IL6RB\_HUMAN&223-311\\
\hline		 	
PF00043&95&6818&4052&6gsu&GSTM1\_RAT&104-192\\
\hline	
PF00044&151&6206&1422&1crw&G3P\_PANVR&1-148\\
\hline	
PF00046&57&7372&1761&2vi6&NANOG\_MOUSE&97-153\\
\hline
PF00056&142&4185&1120&1a5z&LDH\_THEMA&1-140\\
\hline		
PF00059&108&5293&3258&1lit&REG1A\_HUMAN&53-164\\
\hline
PF00071&161&10779&3793&5p21&RASH\_HUMAN&5-165\\
\hline
PF00073&171&9524&487&2r06&POLG\_HRV14&92-299\\
\hline
PF00076&70&21125&10113&1g2e&ELAV4\_HUMAN&48-118\\
\hline				
PF00081&82&3229&890&3bfr& SODM\_YEAST&27-115\\
\hline		
PF00084&56&5831&3453&1elv&C1S\_HUMAN&359-421\\
\hline	
PF00085&104&10569&6137&3gnj&VWF\_HUMAN&1691-1863\\
\hline
PF00091&216&8656&917&2r75&FTSZ\_AQUAE&9-181\\
\hline
PF00092&179&3936&1786&1atz&VWF\_HUMAN&1691-1863\\
\hline	
PF00105&70&2549&816&1gdc&GCR\_RAT&438-507\\
\hline
PF00108&264&6839&2688&3goa&FADA\_SALTY&1-254\\
\hline
\end{tabular}
\caption{Domain families included in our extended study, listed with Pfam ID,
 length $N$, number of sequences $B$ (after removal of duplicate sequences), number of effective
 sequences $B_{eff}$ (under $x=0.8$, i.e., $80\%$ threshold for reweighting), and the PDB and UniProt specifications for the
structure used to access the DCA prediction quality.}
\label{tab:morefamilies}
\end{table*}

\end{document}